\documentclass[10pt,journal,compsoc]{IEEEtran}
\usepackage{booktabs}
\usepackage{xcolor}
\usepackage{url}
\usepackage{subcaption}
\usepackage{graphicx}
\usepackage{amsfonts}
\usepackage{mathtools}
\usepackage{makecell}
\usepackage{hyperref}
\usepackage{multirow}

\hypersetup{
    colorlinks=true,
    linkcolor=blue,
    filecolor=blue,      
    urlcolor=blue,
    }

\DeclarePairedDelimiter\floor{\lfloor}{\rfloor}

\newcommand{\dnnz}{\textit{nnz}}
\newcommand{\flops}{\mathrm{flops}}
\newcommand{\mA}{\mathbf{A}}
\newcommand{\mB}{\mathbf{B}}
\newcommand{\mC}{\mathbf{C}}
\newcommand{\mX}{\mathbf{X}}
\newcommand{\mY}{\mathbf{Y}}
\newcommand{\mx}{\mathbf{x}}
\newcommand{\my}{\mathbf{y}}

\newcommand{\revision}   [1]{{#1}}

\newcommand{\flop}{\mathrm{flops}}
\newcommand{\nnz}{\textit{nnz}}

\newcommand{\spparmat}{{\tt SpParMat}}
\newcommand{\spmat}{{\tt SpMat}}

\begin{document}

\title{Combinatorial BLAS 2.0: Scaling combinatorial algorithms on distributed-memory systems}

\author{Ariful~Azad,
        Oguz~Selvitopi,
        Md~Taufique~Hussain,
        John~R.~Gilbert,
        Ayd\i n~Bulu\c{c}
\IEEEcompsocitemizethanks{\IEEEcompsocthanksitem O. Selvitopi and A. Bulu\c{c} are with Lawrence Berkeley National Laboratory \protect
\IEEEcompsocthanksitem A. Azad and MT. Hussain are with Indiana University \protect 
\IEEEcompsocthanksitem J.R. Gilbert is with University of California, Santa Barbara}
}


\maketitle

\begin{abstract}
  Combinatorial algorithms such as those that arise in graph analysis, modeling of discrete systems, bioinformatics, and chemistry, are often hard to parallelize. 
  The Combinatorial BLAS library implements key computational primitives for rapid development of combinatorial algorithms in distributed-memory systems. 
  During the decade since its first introduction, the Combinatorial BLAS library has evolved and expanded significantly. This paper details many of the key technical features of Combinatorial BLAS 
  version 2.0, such as communication avoidance, hierarchical parallelism via in-node multithreading, accelerator support via GPU kernels, generalized semiring support, implementations of key data structures and functions, and scalable distributed I/O operations for human-readable files.
  Our paper also presents several rules of thumb for choosing the right data structures and functions in Combinatorial BLAS 2.0, under various common application scenarios.  
  \end{abstract}

\section{Introduction}

Combinatorial BLAS, or CombBLAS for short, is a distributed-memory library that provides a set of matrix and vector data structures as well as highly-optimized implementations of fundamental operations on and among those data structures. 
The original purpose of CombBLAS was to provide a proof-of-concept implementation of graph algorithms in the language of linear algebra, demonstrating the feasibility of this approach and the scalability of the resulting implementation. CombBLAS has been used as a benchmark by competing distributed-memory graph libraries. Since launch of the GraphBLAS standardization effort~\cite{mathgraphblas16}, the API development team~\cite{bulucc2017design} relied on the design and naming choices made by CombBLAS. 

Since its inception a decade ago~\cite{combblas2011}, CombBLAS has been used in a wide variety of distributed data analytics and scientific computing applications.  
CombBLAS has also evolved heavily to take advantage of the developments in distributed algorithms and architectures. This paper describes the evolution of CombBLAS over the last decade. The introduction surveys the main contributions; the later sections go into their detail.

\noindent
{\bf Avoiding Communication.}
Communication is the primary bottleneck in scaling data-intensive applications to exascale. Communication-avoiding (CA) algorithms reorganize the computation to reduce communication costs, often asymptotically, and expose more parallelism. CombBLAS was the first library to include a 3D (or 2.5D) sparse matrix-matrix (SpGEMM) multiplication algorithm and since then the algorithm has been expanded to minimize communication under a given (often tight) memory budget. We discuss this integrated CA algorithm in Section~\ref{sec:commavoid}. Recently, we also integrated a 1.5D sparse times tall-skinny dense matrix multiplication (SpMM) algorithm into CombBLAS.

\noindent
{\bf Hierarchical Parallelism.}
Exascale computers are going to be based on either accelerators or multi-core CPUs. Original CombBLAS used to run on a so-called ``flat-MPI'' model where each core was tasked with running an MPI process. With the core counts per compute node increasing from single digits to almost triple digits, a flat MPI model is now known to be unscalable due to increased communication bottlenecks in Network Interface Card (NIC)~\cite{gropp2016modeling}. Several hierarchical programming models have been proposed where the inter-node communication is handled by either MPI or a different distributed communication library, and the intra-node parallelism is handled via a multithreading platform such as OpenMP. It is also possible to use MPI hierarchically where a smaller MPI communicator is used within a node, an approach known as MPI+MPI. 

CombBLAS 2.0 follows the most popular paradigm of using OpenMP parallelism within a node, and MPI for communication across nodes. One reason we avoided a process-based MPI+MPI approach is load imbalance. While CombBLAS avoids most load balance issues by randomly permuting sparse matrices during their assembly, the load imbalance can still hurt the performance if the library runs on 100,000 processes. The use of OpenMP within a node allows CombBLAS to rein in load imbalance since it reduces the number of partitions of a sparse matrix by a factor proportional to the degree of on-node multithreading. \revision{Furthermore, popular sparse data structures such as Compressed Sparse Columns (CSC)~\cite{saad2003iterative} become wasteful as local matrices become hypersparse~\cite{hypersparse08} due to 2D or 3D decomposition on large numbers of partitions. CombBLAS tames hypersparsity either by specialized data structures such as Doubly-Compressed Sparse Columns (DCSC) or by increasing in-node multithreading that results in fewer MPI processes (and hence fewer matrix partitions).} Virtually all CombBLAS functions are multithreaded within a compute node using the state-of-the-art algorithms, the details of which are given in Section~\ref{sec:multithreading}. 

\noindent
{\bf Accelerator Support.}
CombBLAS introduces accelerator support for several key operations to address the shift in supercomputer architectures towards exascale.
Widely used operations with relatively high arithmetic intensity such as SpGEMM, SpMM, or the multiplication of a sparse matrix with a vector (SpMV) can now effectively make use of accelerators. 
Tailored algorithms and optimizations for these key operations aim to utilize all resources of a node through a joint execution framework.
Details about the accelerator support in CombBLAS are described in Section~\ref{sec:gpus}.

\noindent
{\bf User-Defined Operations and Types.}
Several properties set CombBLAS apart from other distributed matrix libraries. The primary one is the capability to use arbitrary functions and semirings in its operations. The ability to pass user-defined functions and semirings allow CombBLAS to be extremely flexible in terms of implementing a wide range of combinatorial algorithms, including those arising in graph computations~\cite{azad2016distributed,azad2017reverse} as well as other areas such as computational biology~\cite{selvitopi2020distributed,guidi2020parallel}. CombBLAS also takes advantage of C++ templates to a high degree, allowing the user to implement previously unspecified operations by just changing the template type of matrix/vector values and the semiring definition. For example, it takes approximately 10 lines of code to perform neighborhood aggregation on vector valued data instead of scalars or to operate on a matrix where each element is another smaller matrix block. 

CombBLAS 1.0 had basic semiring support but it was limited in the sense that the elements of two inputs (matrices or vectors) as well as the output had to come from the same set (e.g., $\mathbb{R}$, $\mathbb{Z}$, or $\mathbb{B}$). However, in a significant portion of graph algorithms, those elements come from different sets. For example, the graph can have floating-point weights whereas the vector indices are a pair of integers~\cite{azad2016distributed}. CombBLAS 2.0 removes this restriction and allows operations on three different types (two for inputs and one for the output), effectively implementing heterogeneous algebras~\cite{birkhoff1970heterogeneous}.

Similarly, CombBLAS 1.0 only allowed one type of index, such as {\tt uint32\_t} or {\tt uint64\_t}. It is often the case that the sparse matrices CombBLAS operates on are too large to be indexed globally with 32-bit integers but they are small enough that the submatrices (or subvectors) assigned to individual processors can be indexed with 32-bit integers. CombBLAS 2.0 allows objects to have two different index types: one for the global indices, say {\tt uint64\_t}, and one of the local indices, say {\tt uint32\_t}. The storage requirements follow the local index size, while the global index is only used for calculating offsets when communicating with other processors or returning global indices to the user, enabling processing of huge datasets that would otherwise be impossible with a single 64-bit index type.

\noindent
{\bf Parallel I/O on Human-Readable Data.}
Ability to input and output data efficiently is a fundamental component of any software library that is used for data-intensive computing. Since CombBLAS is designed for problems that are too large to be efficiently handled by a single compute node, the I/O routines also need to be fully parallelized for them to not be a bottleneck. \revision{CombBLAS 2.0 introduces parallel I/O support for two human readable text formats, in addition to continuing parallel I/O support for our proprietary binary format.}
While processing human-readable text files have their overheads compared to binary formats, I/O has never been a bottleneck in CombBLAS because we have fully parallelized our I/O routines. We explain the details of our parallel I/O in Section~\ref{sec:io}.

\noindent
{\bf Related Work and Availability.}
The GraphBLAS API~\cite{bulucc2017design} is fully implemented by SuiteSparse::GraphBLAS~\cite{davis2019algorithm} for serial and shared-memory executions. GraphBLAST~\cite{yang2019graphblast} also closely follows the GraphBLAS API for GPUs but it takes a more pragmatic approach and only implements a subset of the operations commonly needed for graph algorithms.  
Cyclops Tensor Framework (CTF)~\cite{solomonik2015sparse} is another distributed-memory library that performs matrix computations on arbitrary functions and semirings. 
With the introduction of the GraphBLAS API, we expect more libraries to have the capability of executing on user-defined functions. 
The current repository for Combinatorial BLAS is hosted at \href{https://github.com/PASSIONLab/CombBLAS}{https://github.com/PASSIONLab/CombBLAS}.

\section{Technical Background}
\subsection{Basic Data Structures}
Conceptually at a high level, CombBLAS library has two major data structures: matrices and vectors. Both matrices and vectors have sparse as well as dense versions. Since the primary use case of CombBLAS is data analytics, the overwhelming majority of operations involve at least one sparse matrix. \revision{A comprehensive background on using linear algebra for graph and data analytics tasks is beyond the scope of this paper and we refer the interested readers to the book by Kepner and Gilbert~\cite{kepner2011graph}}.

The data structures are either stored in a single address space (henceforth called \emph{local}) or distributed. We explicitly do not make a distinction between serial and multithreaded when we refer to data structures but the methods that operate on objects are naturally distinguished based on threading. The typical end user is expected to interact with CombBLAS via its distributed classes, e.g., \spparmat\ and {\tt FullyDistSpVec}. 

The distributed sparse matrix class \spparmat\ is parameterized by the abstract base class \spmat, which is a local sparse matrix. Any sparse matrix data structure can derive from \spmat\ as long as it implements the \spmat\ interface, which consists of a few fundamental methods that are needed by CombBLAS. Currently, we have three such classes implemented: {\tt SpDCCols}, {\tt SpCCols}, and {\tt SpTuples}. {\tt SpDCCols} and  {\tt SpCCols} are wrappers around  DCSC and CSC formats, respectively. {\tt SpTuples} is a wrapper around a C++ std::vector composed of std::tuple entries and is mostly used for interfacing with other libraries and for I/O.

\spmat\ requires its derived classes to implement two iterators that allow power users to implement novel graph and data analysis algorithms in a data structure agnostic manner without going through the CombBLAS primary interface. The first iterator iterates over all nonzero columns of \spmat, and the second iterator goes over all the nonzero entries within a column of \spmat. CombBLAS, therefore, favors column-based data structures for higher performance.

\subsection{Data Distribution}
The distinction among 1D, 2D, and 3D distributions is whether the underlying process grid is indexed by one, two, or three components. How exactly the matrix is laid out on a given $n$-dimensional process grid is specific to the underlying distributed object class. CombBLAS does 2D distribution by default for all matrices but it also allows 3D distribution (i.e., distributing into a process grid that is indexed by three components) for its sparse matrices (Section~\ref{sec:caspgemm} and Figure~\ref{fig:data-distribution} describe these distributions). 

For vectors, CombBLAS supports a 2D decomposition by default, which is superimposed over the default 2D matrix decomposition. If a vector $\mathbf{v} \in \mathbb{S}^n$ is distributed on $p$ processes logically organized on a virtual $p_r \times p_c$ grid, then $\mathbf{v_i}$ is the subvector of length $\lvert \mathbf{v_i} \rvert {=}\floor{n/p_r}$ (except for the last piece $\mathbf{v_{p_r{-}1}}$ which is of length $n- (p_r{-}1) \floor{n/p_r}$) that is owned collectively by all the processes in the $i$th process row $P(i,:)$. The subvector $\mathbf{v_i}$ is further partitioned into $p_c$ pieces where each process $P(i,j)$ \emph{exclusively} owns the subvector $\mathbf{v_{ij}}$ of length $\floor{\lvert \mathbf{v_i} \rvert / p_c}$ (same remainder rules apply for parts owned by the last process in a process column, i.e., $P(i,p_c{-}1)$). We emphasize that there is no replication in this vector distribution: to form $\mathbf{v_i}$ on processor $P(i,j)$, an explicit gather operation is needed along process row $P(i,:)$. 


CombBLAS supports two types of distributions for storing dense matrices: a 2D distribution like the one used for sparse matrices and the superimposed 2D distribution used for vectors.
From matrix point of view, the former distribution splits the dense matrix along both rows and columns (except when $p_r{=}1$ or $p_c{=}1$), while the latter distribution splits the dense matrix only along the row dimension.
These two cases have their use-cases in different scenarios: CombBLAS utilizes the superimposed 2D distribution for tall-skinny dense matrices and true 2D distribution for other dense matrices.
All local dense matrices are stored in row-major order.
The choice of distribution for dense matrices leads to different algorithms for operations such as sparse matrix times multiple dense vectors (SpMM) that involve dense matrices.
For example, in parallel distributed SpMM, the superimposed 2D distribution leads a parallel algorithm in which two dense matrices are communicated.
For tall-skinny dense matrices, this is preferable to parallel SpMM algorithms that communicate a sparse and a dense matrix -- which is the case for true 2D distribution~\cite{ics21spmm}.


\subsection{Notation}
The key operations in CombBLAS are among matrices of the kind $\mA \in \mathbb{S}^{m\times n}$ and vectors of the kind $\mathbf{v} \in \mathbb{S}^n$. 
$\mathbb{S}$ is the set of the semiring from which the entries of the matrices and vectors are chosen. 
Both matrices and vectors are distributed on $p$ processors. We use the notation $\mA(i,:)$ to refer to the $i$th row, $\mA(:,j)$ to refer to the $j$th column, and $\mA(i,j)$ to refer to the $(i,j)$th entry of $\mA$. Only $\dnnz(\mA)$ out of $m \cdot n$ possible entries of a sparse matrix are nonzero where $\dnnz(\mA) \ll m \cdot n$.
For notational simplicity in discussing multithreaded algorithms within a node, we denote a local submatrix of $\mA$ by $\tilde{\mA}$.

\section{Reducing Communication}
\label{sec:commavoid}
\subsection{Communication Costs}

While CombBLAS implements many operations, 
there are only a handful of \emph{key operations} that
directly affect the application runtime. Those key operations are computing the product of two sparse matrices (SpGEMM), sparse-matrix dense-vector multiplication (SpMV) and its multi-vector generalization (SpMM), sparse-matrix sparse-vector multiplication (SpMSpV), and vector extraction and assignment (VecAssign). We use the syntax $\mC=\mA \mB$ for SpGEMM, $\my = \mA \mx$ for SpMV and SpMSpV, and  $\mY = \mA \mX$ for SpMM. Here, $\mx$ and $\my$ are (dense or sparse) vectors. $\mA \in \mathbb{S}^{m \times n}$, $\mB \in \mathbb{S}^{n \times k}$, and $\mC \in \mathbb{S}^{m \times k}$ are sparse matrices. $\mX \in \mathbb{S}^{n \times k}$ and $\mY \in \mathbb{S}^{m \times k}$ are dense matrices.

\begin{table}[!t]
    \centering
     \caption{Communication complexity of key CombBLAS operations when matrices and vectors are distributed on a 2D $\sqrt{p}{\times}\sqrt{p}$ process grid. $\flops$ denotes the number of nonzero arithmetic operations required when computing SpGEMM between $\mA$ and $\mB$. For SpMSpV, the density of the input vector is $f$ and the unreduced output vector is $g$. }
    \begin{tabular}{l c c c }
        \toprule 
         Operation & Computation & Latency & Bandwidth \\
         \toprule
         SpGEMM &  $O(\flops/p)$ & $O(\sqrt{p}\lg{p})$ & $O(\dnnz(\mA + \mB)/\sqrt{p})$ \\  
          SpMM & $O(k \cdot \dnnz/p)$  & $O(\lg{p})$ & $O(k(m+n)/\sqrt{p})$  \\ 
         SpMV &  $O(\dnnz/p)$ & $O(\sqrt{p})$ & $O(n\lg{p}/\sqrt{p})$  \\ 
         SpMSpV & $O(\dnnz \cdot f/p)$ & $O(\sqrt{p})$ & $ O((nf+ng)/\sqrt{p})$\\ 
         VecAssign & $O(n/p)$ & $O(p)$ & $ O(n/p)$\\
          \bottomrule
    \end{tabular}
    \label{tab:complexity}
\end{table}

To develop an analytical framework for key CombBLAS operations, we assume that the data and computations are balanced across $p$ processes organized in a $\sqrt{p}{\times}\sqrt{p}$ process grid $P_{2D}$, meaning that each process stores approximately $\dnnz(\mA)/p$ nonzeros of a matrix $\mA$ and $n/p$ entries of a vector of length $n$. We emphasize that this assumption is artificial and just made for simplifying the complexity analysis. CombBLAS allows the user to randomly permute  sparse matrices during matrix assembly, using which load balance can be achieved with high probability~\cite{azad2020distributed}.

Given this balanced setting, Table~\ref{tab:complexity} shows the computation and communication complexity of these operations.
We see that key CombBLAS operations have low computational intensity, and their performances are bound by memory-to-processor and inter-processor communication costs. 
Table~\ref{tab:complexity} also shows that communication becomes increasingly dominant as we move down the table. 
Furthermore, communication latency becomes increasingly significant relative to bandwidth costs as we move from matrix to vector operations. 
This theoretical analysis is also observed in graph algorithms implemented on top of CombBLAS. 
Thus, different algorithms need different strategies to minimize communication depending on the relative importance of computation, bandwidth and latency costs.


\begin{figure*}[t!]
    \centering
    \begin{subfigure}[t]{0.15\textwidth}
        \includegraphics[width=1.0\textwidth]{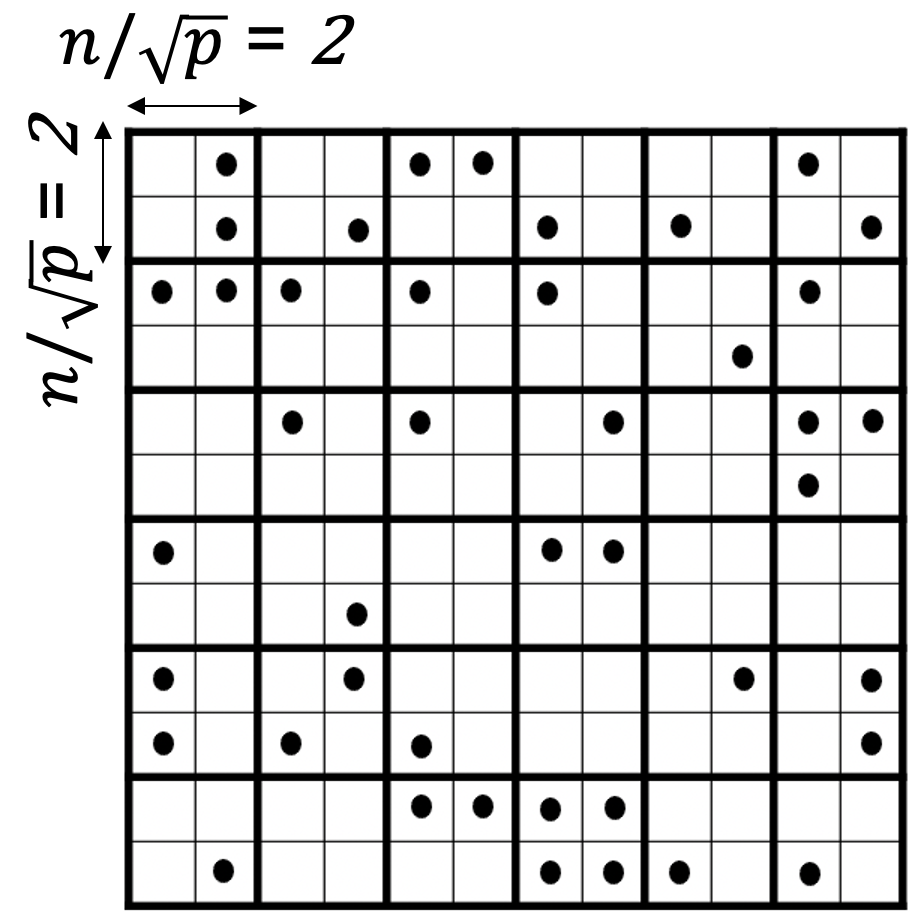}
        \caption{A $12\times12$ sparse matrix distributed in a 2D $6\times6$ grid of $36$ processes.}
        \label{fig:matrix-in-2d}
    \end{subfigure}
    ~
    \begin{subfigure}[t]{0.15\textwidth}
        \includegraphics[width=1.0\textwidth]{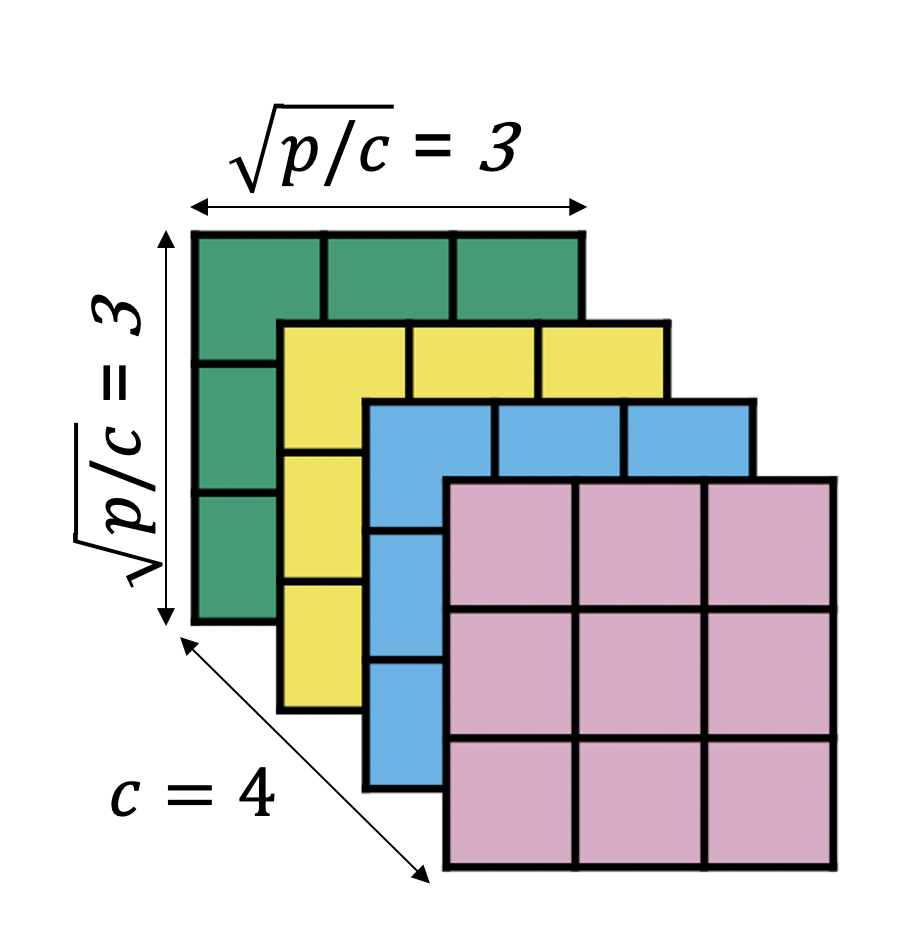}
        \caption{A 3D grid of $36$ processes organized in four 2D $3\times3$ grids.}
        \label{fig:3d-grid}
    \end{subfigure}
    ~
    \begin{subfigure}[t]{0.15\textwidth}
        \includegraphics[width=1.0\textwidth]{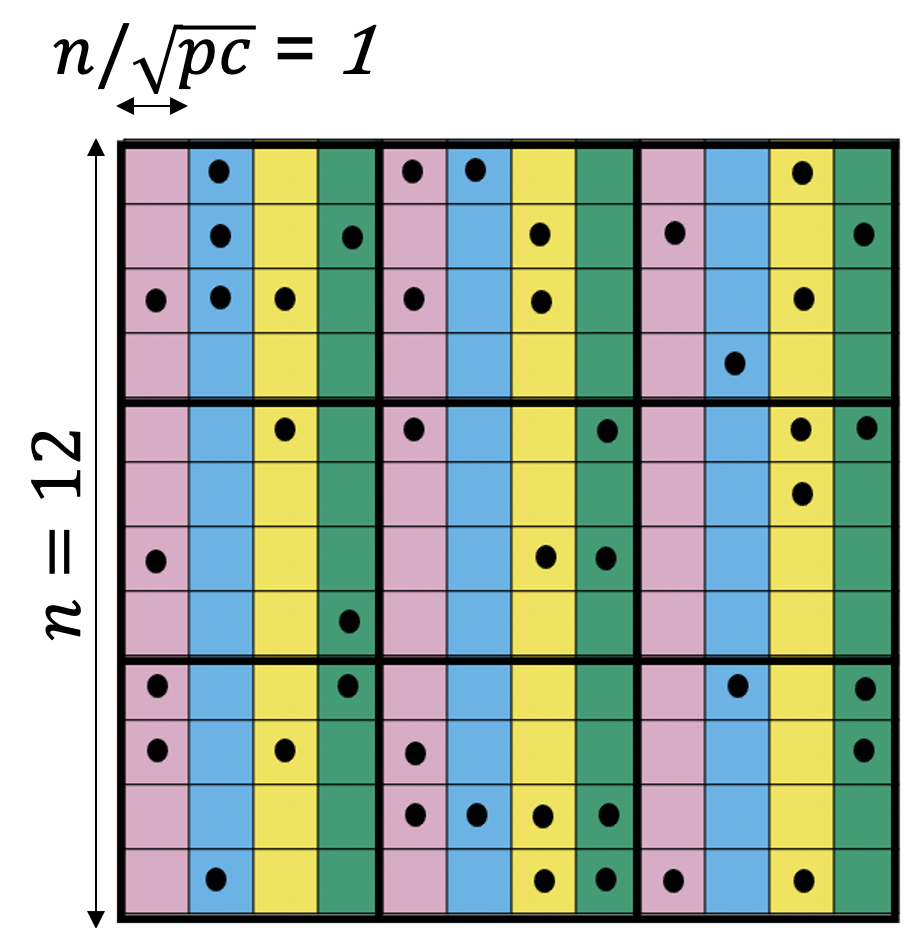}
        \caption{Partitioning $\mA$ into the 3D grid by splitting up the columns.}
        \label{fig:matrix-A-in-3d}
    \end{subfigure}
    ~
    \begin{subfigure}[t]{0.15\textwidth}
        \includegraphics[width=1.0\textwidth]{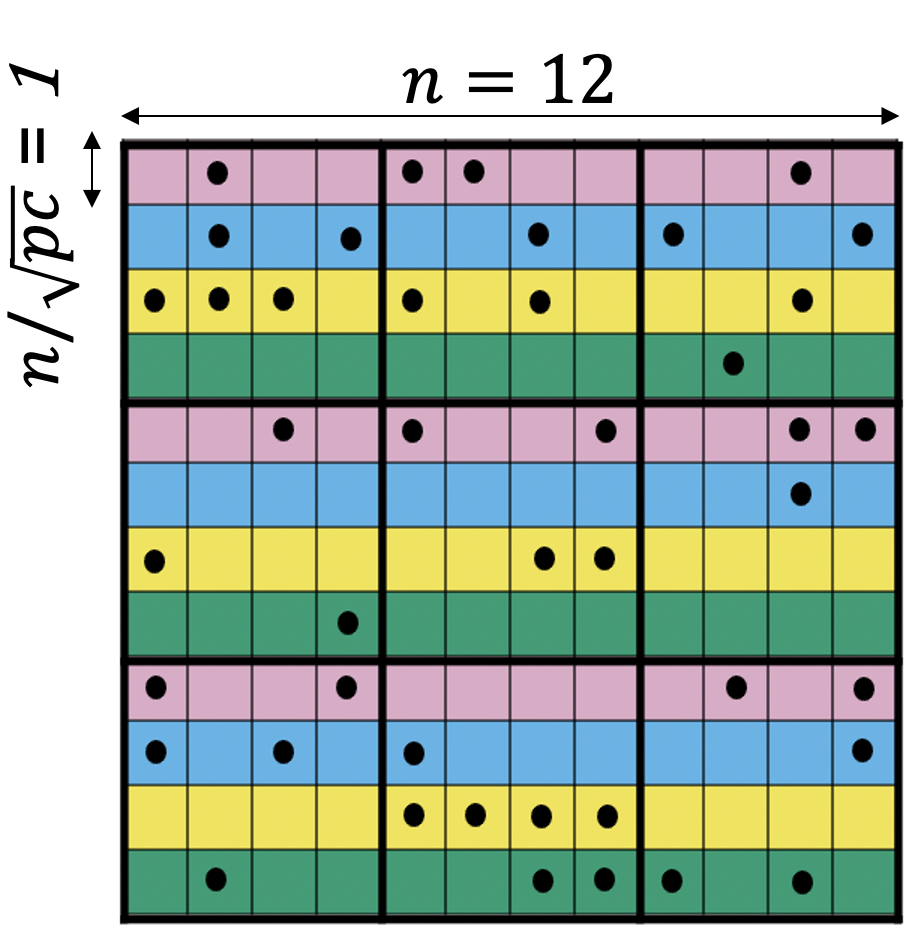}
        \caption{Partitioning $\mB$ into the 3D grid by splitting up the rows.}
        \label{fig:matrix-B-in-3d}
    \end{subfigure}
    ~
    \begin{subfigure}[t]{0.15\textwidth}
        \centering
        \includegraphics[width=1.0\textwidth]{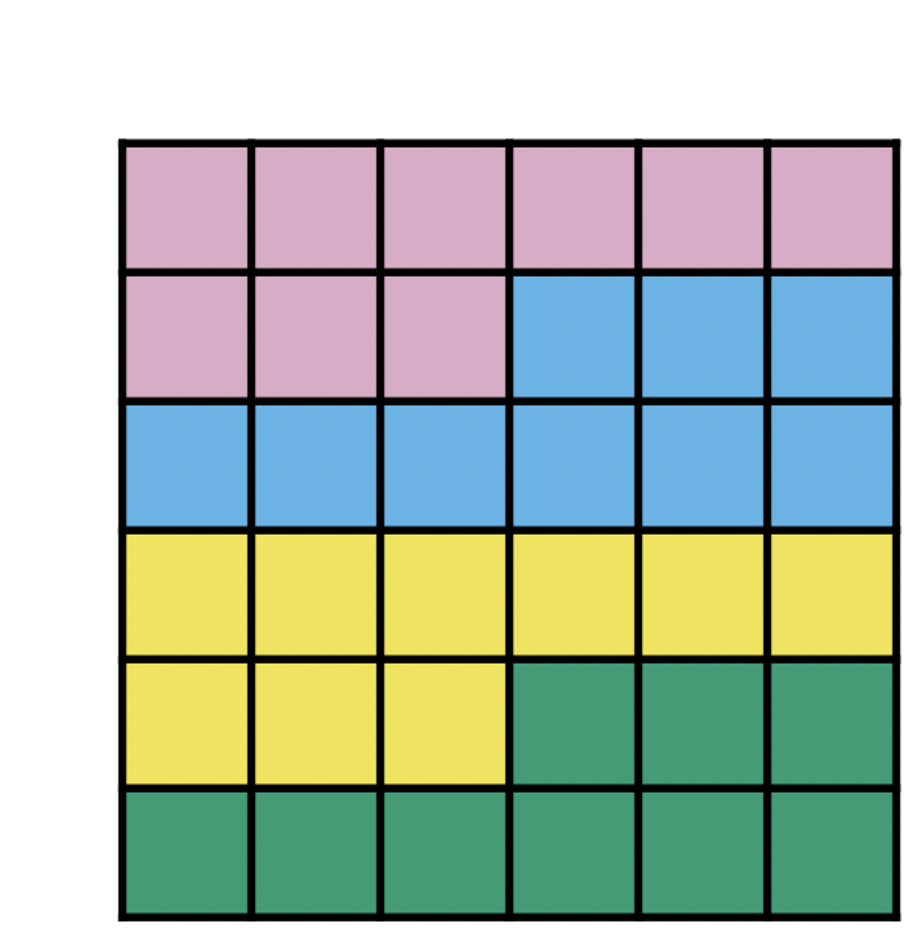}
        \caption{ Converting a $6{\times6}$ grid to a $4{\times}3{\times3}$ grid in the regular way. }
        \label{fig:2d-3d-regular}
    \end{subfigure}
    ~
    \begin{subfigure}[t]{0.15\textwidth}
        \centering
        \includegraphics[width=1.0\textwidth]{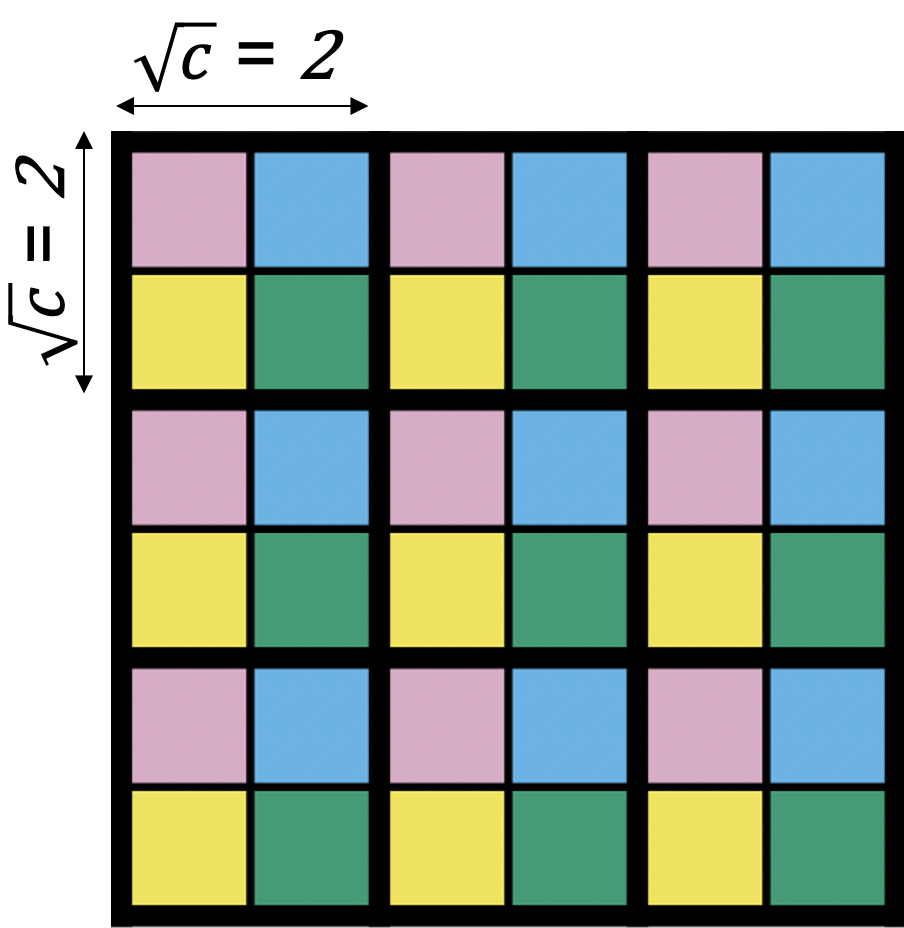}
        \caption{ Conversion from 2D to 3D grid using reduced communicators.}
        \label{fig:2d-3d-special}
    \end{subfigure}
    \caption{2D and 3D distribution of a sparse matrix in CombBLAS. Purple, blue, yellow and green colors represent the first, second, third and fourth layers respectively. 
    }
    \label{fig:data-distribution}
\end{figure*}


\begin{figure}[!t]
    \centering
    \includegraphics[width=0.45\textwidth]{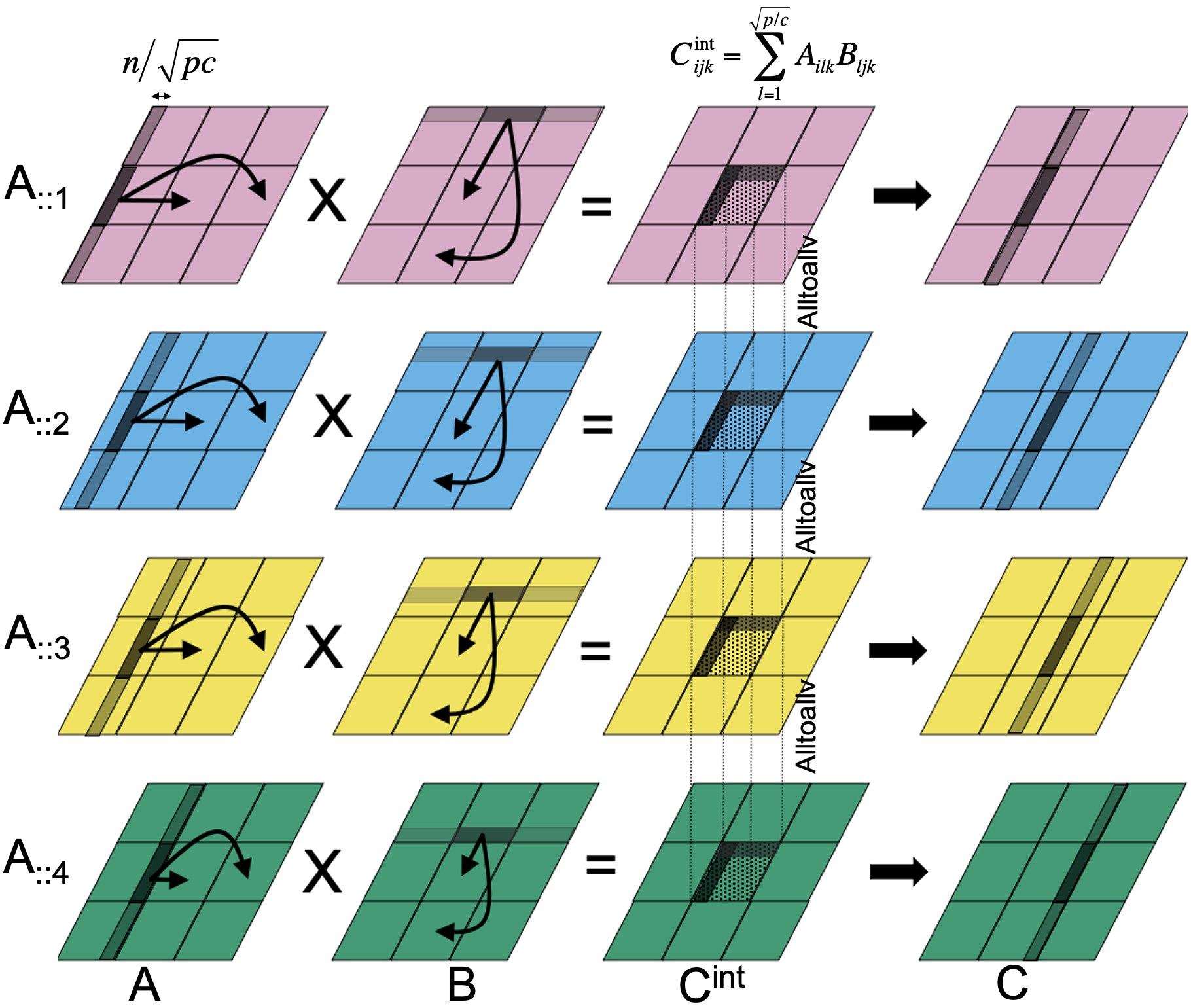}
    \caption{ Execution of the communication avoiding SpGEMM for multiplying sparse matrices $\textbf{A}$ and $\textbf{B}$ to get $\textbf{C}$ on a $c \times \sqrt{p/c} \times \sqrt{p/c}$ process grid. Shown operations are involved to generate the local portion of $\textbf{C}$ only for the processes at the second row and the second column of each layer during the first stage of the algorithm. 
    }
    \label{fig:summa3d}
\end{figure}

\revision{
Reducing communication cost can be achieved either algorithmically, which provides asymptotic improvements, or via methods that change implementation details (in particular, managing fine-grained communication) to reduce costs without changing the asymptotic complexity. The former only applies to operations that have arithmetic intensity bounded above 1. Hence, we apply asymptotically better communication-avoiding methods to SpGEMM and various fine-grained communication schemes to other operations. 
}

\subsection{Communication-avoiding SpGEMM}
\label{sec:caspgemm}
In earlier versions of CombBLAS, distributed SpGEMM was based on the 2D SUMMA algorithm~\cite{Geijn1995} adapted for sparse matrices~\cite{Buluc2012}.
In 2D Sparse SUMMA, input matrices $\mA$ and $\mB$ and the output matrix $\mC$ are distributed on a $\sqrt{p}\times \sqrt{p}$ process grid.
The simplest variant of 2D Sparse SUMMA operates in $\sqrt{p}$ stages.
In the $i$th stage, each process in the $i$th process column
broadcasts their local piece of $\mA$ horizontally (along the process row) and each process in the $i$th process row broadcasts their local piece of $\mB$ vertically (along the process column).
The received submatrices are locally multiplied by each process using a multithreaded SpGEMM algorithm (Section~\ref{sec:multithreading}).
After $\sqrt{p}$ stages, partial results from all stages are merged to obtain the final result. 

The costs of broadcasting input matrices in 2D SUMMA quickly become a performance bottleneck at extreme scale~\cite{Buluc2012, 3dspgemmsisc16}. 
To alleviate this bottleneck, CombBLAS 2.0 includes CA SpGEMM algorithms, following the success of CA algorithms in dense linear algebra~\cite{Ballard2011}. 
Our CA SpGEMM algorithms distribute matrices on a $c\times \sqrt{p/c}\times \sqrt{p/c}$ process grid, where $c$ denotes the number of layers in the third dimension. 
Fig.~\ref{fig:3d-grid} shows an example of a 3D process grid with four layers, where  each layer is equivalent to a $\sqrt{p/c}\times \sqrt{p/c}$ 2D process grid. 
To facilitate 2D SUMMA algorithm in each layer, we split $\mA$ along the column and $\mB$ along the row into $c$ pieces and then distribute different pieces to different layers as illustrated in Figs.~\ref{fig:matrix-A-in-3d} and \ref{fig:matrix-B-in-3d}.

After input matrices are distributed on a 3D process grid, each layer runs an instance of the 2D SUMMA algorithm to obtain intermediate per-layer results $\mC^{int}$  as shown in Fig.~\ref{fig:summa3d}. 
Here, each layer broadcasts submatrices of $\mA$ along the process rows and submatrices of $\mB$ along the process columns on the 2D grid represented by the layer.
Since these broadcasts materialize on a smaller (by a factor of $\sqrt{c}$) communicator, their costs are reduced at extreme scale~\cite{3dspgemmsisc16}.
After each layer completes their 2D multiplications, the partial results are communicated across layers via an Alltoall communication.
We form the final result $\mC$ by merging pieces received from all layers. 
Since $\mA$ and $\mB$ are distributed differently on the 3D grid, we distribute $\mC$ like $\mA$ (as shown Fig.~\ref{fig:summa3d}).

\noindent
{\bf Guideline on selecting the number of layers ($c$).} 
Unlike dense CA algorithms~\cite{Ballard2011} that replicate input matrices to reduce communication, our CA SpGEMM splits input matrices and does not require any extra memory for inputs with increasing number of layers. 
Generally, the time required to broadcast $\mA$ and $\mB$ decreases as we increase $c$ (e.g., we could completely eliminate broadcasts by using an $p\times 1\times 1$ grid).
However, as $c$ increases, the costs of inter-layer Alltoall communication and the final merging also increase.
Furthermore, the memory required to store intermediate results increases with increasing number of layers.
Therefore, it is challenging to find the optimum $c$ as it depends on the tradeoff between broadcasts and inter-layer Alltoall communication costs, as well as the available memory. 
Our general guideline is to select $c$ with $c \leq \sqrt[3]{p}$ so that inter-layer Alltoall does not dominate intra-layer broadcasts.


\noindent
{\bf Conversion between 2D and 3D distributions}. 
At present, CombBLAS performs I/O only with 2D matrices (Section~\ref{sec:io}).
In order to use CA SpGEMM algorithms, we convert matrices from a 
$\sqrt{p} {\times} \sqrt{p}$ process grid to a $c {\times}\sqrt{p/c} {\times} \sqrt{p/c}$ grid. 
\revision{
Fig.~\ref{fig:matrix-in-2d} and  Fig.~\ref{fig:3d-grid} show an example how a $12{\times}12$ sparse matrix is 
converted from a $6{\times}6$ grid to a $4{\times}3{\times}3$ grid. CombBLAS provides two ways to create 3D matrices from 2D matrices.}
 \revision{
Fig.~\ref{fig:2d-3d-regular} shows how the conversion is done in the regular way where processes on a 2D grid are numbered in the row-major order and we place $p/c$ processes numbered $\{i,i{+}1,...,i{+}p/c{-}1\}$ into the $i$th layer. 
This conversion redistributes a 2D matrix on a 3D grid using an Alltoallv operation among all processes.
In the second approach, we reinterpret the whole 2D grid as a $\sqrt{p/c} \times \sqrt{p/c}$ supergrid (shown with thick lines in Fig.~\ref{fig:2d-3d-special}) where each cell of the supergrid has a $\sqrt{c}{\times}\sqrt{c}$ subgrid of the 2D grid. 
Thus, each supergrid cell corresponds to one cell in each layer of the 3D grid. 
We then assign each cell of a subgrid to the corresponding cell in each layer of the 3D grid as shown in Fig.~\ref{fig:2d-3d-special}.  
Here, $p/c$ Alltoallv calls run in parallel with each Alltoallv involving $c$ processes.
By operating on a smaller communicator, the latter approach 
reduces the communication cost at extreme scale. 
For example, when converting the Metaclust50 matrix from a $64{\times} 64$ grid to a $16{\times} 16 {\times} 16$ grid on 256 nodes, the second approach takes 14 seconds whereas the first approach takes 18 seconds. }


\subsection{\revision{Fine-grained communication-reduction schemes}}
\revision{For operations with lower arithmetic intensities,  CombBLAS provides several other communication-reduction schemes that are often tailored for different applications.}
We broadly group them into three classes and discuss them in relation to their corresponding applications.

\noindent
{\bf Vector replications to avoid asynchronous communication in graph matching algorithms.}
\revision{CombBLAS powers distributed-memory maximal cardinality matching~\cite{maxilmalparco16}, 
maximum cardinality matching~\cite{azad2016distributed} and heavy-weight perfect matching~\cite{azad2020distributed} algorithms for bipartite graphs.}
All of these algorithms iteratively match vertices between two parts of a bipartite graph, where endpoints of currently matched edges
are stored in two dense vectors (instances of {\tt FullyDistVec}).
In iterative matching algorithms, these dense vectors are accessed irregularly to discover alternating paths (paths with alternating matched and unmatched edges) and then updated when the matching is improved. 
However, {\tt FullyDistVec} does not perform well for 
such irregular and asynchronous accesses or updates because it  distributes all $n$ entries of the vector uniformly among $p$ processes. 
To alleviate this problem, we replicated dense vectors within a process row and a process column so that each process stores $n/\sqrt{p}$ entries of the vector as opposed to $n/p$ entries stored in the default distribution. 
Such a vector replication scheme eliminates fine-grained communication among processes and promotes updates in the bulk synchronous fashion. 
Since the vector is replicated $\sqrt{p}$ times, the memory overhead is significant only when $\sqrt{p} \gg d$, where $d$ is the average degree of the graph.

\noindent
{\bf One-sided communication}
Even though CombBLAS performs the best with bulk-synchronous parallel executions, it also utilizes asynchronous one-sided communication if needed by certain algorithms. 
\revision{For example, it is inefficient to traverse long paths level-by-level in the bulk-synchronous fashion because of limited parallelism and high synchronization overheads.
For this case, we asynchronously traverse long paths using special hardware features such as Remote Memory Access operations (e.g., one-sided put and get operations in MPI).
Currently, this approach is only implemented in maximum-cardinality matching algorithms~\cite{azad2016distributed}, but it could potentially be used for SpMSpV as well.}
 

\noindent
{\bf Collective communication on reduced communicators.}
Irregular operations performed solely on sparse/dense vectors are harder to scale because of the performance bottleneck arisen from the communication latency that increases linearly with $p$ (see  Table~\ref{tab:complexity}).
This problem exacerbates for iterative algorithms with skewed communication patterns where the data sent and received by processors changes dramatically over iterations. 
\revision{For example, different iterations of the distributed Awerbuch-Shiloach algorithm~\cite{laccIPDPS19} for finding connected components often process drastically different numbers of vertices. This results in highly-skewed communication where different processes exchange different volumes of data. 
In the presence of such skewed communication, vector assignment or extract operation implemented with the Alltoall collective executed over the entire process grid does not perform well in practice. 
CombBLAS alleviates this problem by providing assign/extract operations in time proportional (or as close to possible) to the size of the updated or extracted region.
To this end, we identify a small set of processes that send or receive more than 90\% of the messages and process them separately with broadcast or reduction operations. 
The rest of the processes then create a reduced sub-communicator 
where another Alltoall takes place. 
Currently, this approach is implemented only for vector operations inside the connected component algorithm. 
However, it can be easily extended to SpMSpV and other vector operations as well.}

\section{Node-level multithreading}
\label{sec:multithreading}

\begin{table*}[!t]
    \centering
    \caption{Multithreaded algorithms implemented in CombBLAS 2.0. }
    \begin{tabular}{l l l}
        \toprule 
         Operation & Work per thread & Algorithms \\
         \toprule
         SpGEMM~\cite{3dspgemmsisc16, hashspgemmparco19} & a subset of  columns in the output matrix &  column-by-column using heap, hash table, or hybrid heap-hash\\ 
          SpMM & a subset of  columns in the input matrix &  \\
         SpMV & a subset of  columns or rows in the input matrix &  inner product\\ 
         SpMSpV~\cite{spmspvipdps17} & a subset of nonzeros  in the input vector & merging using SPA, heap, sorting, or binning\\ 
          \bottomrule
    \end{tabular}
    
    \label{tab:multithreading-options}
\end{table*}

Hybrid parallelization using OpenMP/pthread for in-node computation and MPI for inter-node communication has become the dominant programming model for applications running on modern supercomputers. 
This hybrid parallelization comes with two apparent benefits as it avoids data replication among threads within a process and reduces the number of messages exchanged among processes. 
For various sparse computations in CombBLAS, the hybrid parallelism provides three other crucial benefits.
If the number of processes reduces from in-node multithreading, (1) the sparse SUMMA algorithm for SpGEMM may need less memory to store intermediate results, 
(2) the communication cost of latency-bound operations such as SpMSpV reduces, and 
(3) irregular pointer chasing operations such as vector assign become less expensive. 
Hence, CombBLAS 2.0 includes multithreaded algorithms for almost all operations.
For SpGEMM, SpMV, and SpMSpV, we developed multiple algorithms because different algorithms perform better for matrices with different sparsity patterns.
SpMM has only one variant due to its lower popularity among those operations.
Table~\ref{tab:multithreading-options} summarizes algorithmic options available for various operations in CombBLAS 2.0.
All multithreaded algorithms are implemented using OpenMP. 


\subsection{Multithreaded SpGEMM} 
In-node SpGEMM in CombBLAS is parallelized based on Gustavson’s algorithm~\cite{gustavson1978two} that constructs columns of the output matrix in parallel.
To compute the $j$th column $\tilde{\mC}(:,j)$ of the output, Gustavson's algorithm merges columns of $\tilde{\mA}$ corresponding to nonzeros in $\tilde{\mB}(:,j)$. 
CombBLAS 2.0 provided one of the first multithreaded SpGEMM algorithms based on heap~\cite{3dspgemmsisc16} and hash
table~\cite{hashspgemmparco19} data structures. 
SpGEMM implementations in CombBLAS operate in three phases. (1) Estimate the total $\flops$ needed by the entire multiplication and then divide columns of the output across threads so that each thread performs  approximately equal work. The complexity of this step is  $O(nnz(\tilde{\mB}))$ when both $\tilde{\mA}$ and $\tilde{\mB}$ are stored in CSC or DCSC formats. 
(2) A symbolic phase estimates the required memory for the output. This step uses the heap or hash table data structure to compute the pattern of $\tilde{\mC}$, and its complexity is $O(\flops)$.  
(3) Finally, the computational phase performs the actual multiplication using the heap or hash table data structure, which has the same $O(\flops)$ complexity as the previous stage. 
We refer to the SpGEMM that uses heap and hash data structures as heap-SpGEMM and hash-SpGEMM, respectively

The heap-SpGEMM algorithm usually performs better than hash-SpGEMM when the SpGEMM has a low compression ratio (the ratio of $\flops$ to nonzeros in the output).
By contrast, hash-SpGEMM excels when the SpGEMM has a high compression ratio~\cite{hashspgemmparco19}. 
We also implemented a hybrid hash-heap algorithm that uses either a heap or a hash table to form a column of $\tilde{\mC}$ depending on the compression ratio of the individual column.
This fine-grained decision helps when matrices have skewed and unpredictable nonzero distribution.  

\subsection{Multithreaded SpMV}
CombBLAS 2.0 includes two variants of multithreaded SpMV. Both of them access the matrix column-by-column due to the DCSC and CSC formats used for storing local matrices.  
In the first variant, the input matrix $\tilde{\mA}$ is partitioned row-wise into $t$ parts where each part $\tilde{\mA}_i$ with $n/t$ rows is stored in the DCSC or CSC format.
Then, each thread multiplies one part $\tilde{\mA}_i$  of $\tilde{\mA}$ by the input vector and writes to a part of the output vector corresponding to the rows in $\tilde{\mA}_i$.
In the second variant, $\tilde{\mA}$ is partitioned column-wise into $t$ parts where each part $\tilde{\mA}_i$ has $n/t$ columns. 
Then, each thread multiplies $\tilde{\mA}_i$  by a part of the input vector corresponding to the columns in $\tilde{\mA}_i$ and writes the output to a thread-private output vector $y_i$.
At the end of the multiplication, all thread-private outputs are accumulated to form the final vector in parallel by all threads. 

Both row- and column-partitioned algorithms have their advantages and disadvantages. 
The row-partitioned SpMV does not require extra memory to store partial results and enjoys better locality in accessing the output vector, but the entire input vector is accessed by all threads. Furthermore, row-wise partitioning of DCSC and CSC matrices incurs significant overhead that can only be mitigated in interactive algorithms. 
In contrast, the column-partitioned SpMV does not need a partitioning step and each thread accesses only a part of the input vector, but it needs extra thread-private memory to store partial results.
While users can use either of these algorithms, the row-partitioned SpMV is recommended in iterative algorithms and the column-partitioned SpMV is recommended for non-iterative algorithms. 


\begin{figure}[!t]
    \centering
    \includegraphics[width=0.95\linewidth]{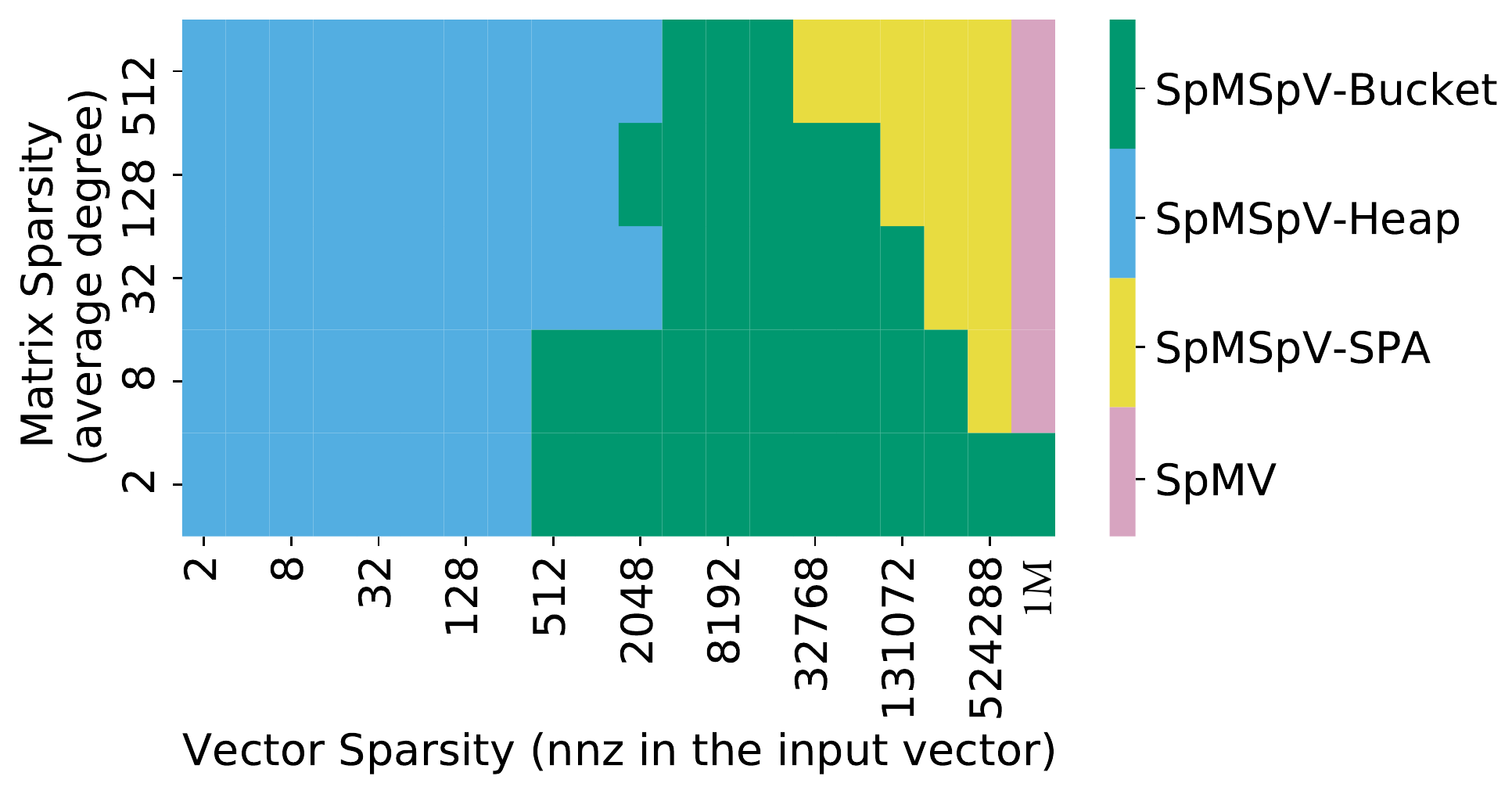}
    \caption{Multiplying an 1M$\times$1M R-MAT sparse matrix by a sparse vector. We show best performing SpMSpV or SpMV algorithms for different matrix and vector sparsity patterns.
    Experiments were run on a single node of the Cori Haswell partition with 32 cores. 
    }
    \label{fig:randSpMV}
\end{figure}

\subsection{Multithreaded SpMSpV}
Given a local sparse matrix $\tilde{\mA}$ and a local sparse vector $\mx$ with $f$ nonzeros, an SpMSpV computes a potentially sparse vector $\my \gets \tilde{\mA} \mx$.
Similar to SpMV, the implementations of SpMSpV in CombBLAS rely on row-wise and column-wise partitioning of the matrix by splitting it into $n/t\times n$ and $n\times n/t$ submatrices for row-wise and column-wise partitions, respectively.
For both partitioning schemes, we implemented several multithreaded SpMSpV algorithms that use different strategies to merge columns of the input matrix corresponding to nonzero entries of the input vector.
\revision{
For the row-partitioned case, CombBLAS 2.0 uses heap (SpMSpV-Heap) and SPA (SpMSpV-SPA) to merge columns within each partition. 
By contrast, the column-partitioned algorithms use buckets for merging (SpMSpV-Bucket).}

To understand algorithmic variants of SpMSpV, let $d$ be the average $\nnz$ per column of $\tilde{\mA}$ and $t$ be the number of threads.
Then, the asymptotic lower bound of SpMSpV is $\Omega(df)$.
Each thread in a row-partitioned SpMSpV needs to access the entire input vector, giving us the parallel complexity of $O(f+df/t)$.
This results in work-inefficient algorithms when $t{>}d$.
By contrast, a column-partitioned algorithm partitions $\mx$ among threads, but needs synchronization when updating the output. 
To alleviate thread synchronizations, we used intermediate buckets to block $\my$ along the rows~\cite{spmspvipdps17} (also known as propagation blocking~\cite{beamer2017reducing}).
Section ~\ref{sec:perfSpMV} shows that different SpMSpV algorithms excel for different sparsity patterns of matrices and vectors. 

\subsection{Multithreaded SpMM}
CombBLAS 2.0 includes only one variant for multithreaded SpMM, because it not as frequently used as the operations discussed so far.
The multithreaded algorithm used for SpMM is similar to the ones used for SpMV and SpMSpV where the sparse matrix is partitioned column-wise.
The dense matrices are stored in row-major order.
These choices result in the preference of reuse on the input dense matrix.
The overhead due to row-wise partitioning of the sparse matrix may be more readily justified in SpMM than in SpMV or SpMSpV, prompting development of such alternative implementations in future CombBLAS releases.
%

\subsection{Performance of SpMSpV and SpMV algorithms with respect to matrix and vector sparsity}
\label{sec:perfSpMV}
As discussed above, CombBLAS 2.0 includes a rich collection of multithreaded SpMSpV and SpMV algorithms.
Different algorithms excel with different sparsity patterns of the input matrix and the vector. 
\revision{To provide a guideline for users, we
used a $1\text{M}\times 1\text{M}$ randomly-generated R-MAT matrix~\cite{chakrabarti2004r}, which is a synthetic graph generator used by the Graph500 benchmark. We used the same parameters as the Graph500 benchmark.
We multiplied the R-MAT matrix
by a vector on a 32-core Intel Haswell processor.}
We change the sparsity of the matrix by changing the average number of nonzeros per column in the matrix generator. 
We also control the sparsity of the input vector by filling it with nonzeros at  uniformly random indices. 
\revision{
Fig.~\ref{fig:randSpMV} summarizes the best performing algorithm for different matrix and vector sparsity patterns. 
We observe that SpMSpV-Heap performs the best with vectors densities less than 0.5\%, SpMSpV-Bucket performs the best with vectors densities greater than 0.5\% and less than 10\%, and  SpMSpV-SPA performs the best with denser matrix and vector.
Interestingly, SpMSpV-SPA and SpMSpV-Bucket perform similarly to or better than our SpMV implementations even when the vector is 50\% dense.
While there is room to improve the performance of  our SpMV implementation, these results suggest that our SpMSpV implementations attain good performance for almost all sparsity levels in the input vector.
}


\section{GPU acceleration}
\label{sec:gpus}


{\bf Operation support.}
CombBLAS 2.0 supports NVIDIA GPU accelerators for relatively more compute-intensive and important operations. 
The supported three such operations are SpGEMM, SpMM, and SpMV.
Although the computations on sparse matrices are often ill-suited for accelerators due to their low computational density and irregular memory access patterns, the cuSPARSE library\footnote{https://docs.nvidia.com/cuda/cusparse/index.html} often provides satisfactory performance compared to its counterpart sparse libraries on CPUs, such as MKL, even for SpMV, which has the lowest arithmetic intensity among the three operations.
A relatively modest speedup of $3\times$ over the CPU often translates into valuable computational improvement and justifies the overhead of copying data back and forth between the host and the device.
The SpGEMM and SpMM operations offer even greater promise for relying on accelerators.
Moreover, recent works~\cite{Nagasaka2017,Gale2020} in this area report $2{-}4\times$ speedups over cuSPARSE, making GPUs even more viable for these operations if they are adapted.
Based on these arguments, CombBLAS 2.0 provides support for these three operations and certain tailored optimizations.
We next discuss how CombBLAS 2.0 handles these operations.

The above-mentioned operations are rarely used alone and they are often part of a larger execution framework that possibly contains other sparse matrix and vector operations.
Since such operations in CombBLAS 2.0 are only realized for CPUs and a big proportion of them would make little sense to offload to GPUs, the matrix and vector objects in GPU-enabled CombBLAS 2.0 still dwell in the host memory.
One main reason for this choice is related to the nature of the graph algorithms.
Most of the graph algorithms require custom binary operators, semirings, and/or custom vector/matrix elements.
Support for such customization on GPUs is nearly nonexistent.
There are efforts such as cusp\footnote{https://github.com/cusplibrary}, but the support is limited and the library is poorly maintained and  occasionally crashes.
Hence, CombBLAS 2.0 assumes that its objects mainly live on the host memory and utilize accelerators on a need basis where there is a promise of computational improvement.
It has the capability for different processes in the SpGEMM, SpMM, or SpMV to utilize CPU or GPU based on the flops of the target operation.
Note that the GPUs are only utilized for the arithmetic semiring.
In the existence of non-built-in types for vector/elements or other semirings, CombBLAS 2.0 falls back to CPUs.

\begin{figure}[t]
\centering
\includegraphics[width=1.0\columnwidth]{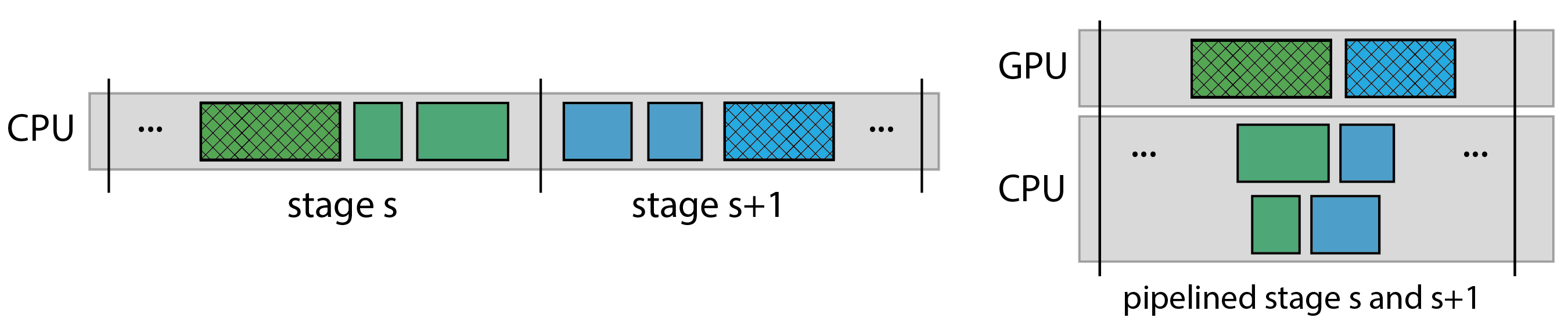}
  \caption{CPU-GPU joint execution of multi-stage parallel algorithms. The rectangles illustrate various tasks and the patterned rectangles illustrate local computations offloaded to  GPU. Any task between two local computations is handled by the host and overlapped with GPU computations. 
  }
  \label{fig:cpu-gpu}
\end{figure}

\noindent
{\bf Coordination.}
The distributed algorithms for SpGEMM, SpMM, and SpMV can be distinguished into two according to whether the local computations are split across multiple stages that are separated by various communication operations and other tasks.
The parallel algorithms that split computations include the Sparse SUMMA algorithm for SpGEMM and all the SpMM algorithms where the dense matrices are 2D partitioned.
The algorithms for SpMV and the SpMM algorithms in which the dense matrices are 1D partitioned do not split computations.
For the former case, we adopt a strategy where the device is responsible only for the local computations and the host is responsible for the rest of the tasks.
%
%
%
In this direction, we aim to overlap the local computations on GPU in stage $s$ with CPU tasks in stage $s$ or $s+1$ (Fig.~\ref{fig:cpu-gpu}).
Moreover, we distribute CPU tasks that do not depend on each other among the CPU cores of the node.
This approach results in pipelined versions of these multi-stage parallel algorithms where the host and the device share work.
%

CombBLAS 2.0 can utilize multiple devices on a node via node-level parallelism.
The work distribution among multiple devices on a node depends on the operation.
For SpGEMM, among the two local input matrices to be multiplied, the first one is replicated among all devices and the second input matrix is partitioned column-wise among the GPUs.
This partitioning conforms to the sparse matrix storage formats and aims to assign roughly equal number of nonzeros to each device.
Each device computes a column stripe of the output matrix, which is concatenated with other stripes at the end.
For SpMM and SpMV, the sparse matrix is column-wise partitioned in a load-balanced manner among the GPUs along with a conformal distribution of the required input dense matrix/vector.
Each device computes partial results for the output matrix/vector which are then reduced by the host.
CombBLAS 2.0 uses threads on the host side to manage GPU coordination and any necessary pre-/post-processing.

\noindent
{\bf Optimizations.}
Among the three supported operations on GPUs, CombBLAS 2.0 has comprehensive optimization for SpGEMM.
The ubiquitous nature of this operation together with its high-level complexity necessitate algorithmic techniques for efficient execution on systems with accelerators.
For this operation, CombBLAS 2.0 makes use of multiple SpGEMM libraries~\cite{Nagasaka2017, Gremse2018, Liu2014} on GPUs by selecting the most appropriate library according to factors such as compression factor and flops.
Other optimizations include faster memory requirement estimation that involves approximate algorithms and a binary merge scheme that spreads out the computations related to merging of the partial results across stages of the SUMMA algorithm.
A recent work~\cite{selvitopi2020optimizing} contains more information about these optimizations.

\section{High-Performance Parallel I/O}
\label{sec:io}
I/O has increasingly become a bottleneck on exascale systems. The graph community does not have a commonly accepted binary exchange format. \revision{CombBLAS defined its binary format since version 1.0 and provided conversion routines. In addition to continuing the support for binary I/O}, CombBLAS 2.0 introduces support for parallel I/O
on two types of human-readable text file formats.

First is the Matrix Market format of NIST (i.e., .mtx files)~\cite{boisvert1996matrix}.
The Coordinate Format of Matrix Market Exchange Format 
is composed of a header telling whether the matrix is symmetric and the type of nonzero values (\emph{real}, \emph{integer}, \emph{pattern}), followed by a tab separated line giving row ($M$), column ($N$), and nonzero counts ($\dnnz$). The body of the text has as many lines as the number of nonzeros in the matrix, with each line listing the row id, column id, and value of one nonzero. The row and column ids are integers in range $\{1,\ldots,M\}$ and $\{1,\ldots,N\}$, respectively.

The second is the label format of the original Markov Clustering implementation~\cite{Dongen2000}. This format is similar to the matrix market format except that (a) the header is not present, and (b) the row and column ids can be arbitrary strings. This label format is convenient because it does not require row and column ids to be in consecutive $\{0,....,M{-}1\}$ and $\{0,....,N{-}1\}$ ranges. It can handle scattered integers in a wide range by treating them as arbitrary labels. The labels do not even need be integers, e.g., the reader can handle labels that can represent DNA or protein sequences. 

The input reader for the Matrix Market case ({\tt ParallelReadMM} function) is relatively straightforward where processor $p_i \in P$ moves its cursor to location $\mathit{filesize} \cdot i/ \vert P \vert$ in the file. It then fast forwards until it sees a newline to start reading. Conversely, if a process' share of the input ends in the middle of a line, that process is responsible for reading the line until completion. The actual reading happens using MPI I/O collective routines into an in-memory byte stream, which is then parsed locally in each processor. In order to write a Matrix Market file, each process simply writes its set of local nonzeros into a byte stream. The first process ($p_o$) creates the file and writes the header information. The writing of the body of the .mtx file happens by writing the byte streams into disk using collective MPI I/O operations. This ensures a human readable text output that is also written in parallel. 

The case of the label format is more complicated. The file reader needs to internally map arbitrary labels into a consecutive range of integers. This is accomplished by a two pass algorithm in our implementation that is called {\tt ReadGeneralizedTuples}. In the first pass, row (and column) ids are hashed randomly into the range $\{0,\ldots, \mathit{max} \}$. That range itself is partitioned into $\vert P \vert$ buckets, one per processor. Then an \emph{all-to-all exchange} sends the row (and column) ids as well as their hashed values to their corresponding processes determined by the bucketing based on hash values. The recipients remove duplicates using a local set data structure. The sizes of the resulting local sets are used to find starting id offsets in each process via an \emph{exclusive scan}, using which each process gives a unique consecutive id to the rows and columns. Each recipient also keeps track of the data sent in this \emph{all-to-all exchange} so it can send back the new ids, which are in the ranges $\{0,\ldots,M{-}1\}$ and $\{0,\ldots,N{-}1\}$, back to each sender. The original senders can now build a string-to-integer map that they can use to relabel their arbitrary identifiers. The two-pass algorithm does not need to go back to disk for the second pass if there is enough memory to cache the first read, but our default implementation re-reads from the disk for a streaming relabeling operation that is more memory frugal. For a file that contains $O(\dnnz)$ nonzeros, the per-process memory consumption of  \texttt{ReadGeneralizedTuples} is $O(\dnnz / \vert P \vert)$ in expectation. The input is already deterministically partitioned into $\vert P \vert$ equal parts and the random hash-based assignment ensures load balance in expectation.

Upon completion, \texttt{ReadGeneralizedTuples} returns two objects: (1) a CombBLAS compliant distributed sparse matrix object, and (2) a CombBLAS compliant distributed vector that maps the newly created integer labels $\{0,\ldots,M{-}1\}$ and $\{0,\ldots,N{-}1\}$ into their original string labels so that the program can convert the internal labels back into their original labels for subsequent processing or while writing the output. A crucial positive side effect of the \texttt{ReadGeneralizedTuples} function is that it automatically permutes row and column ids randomly during the relabeling, ensuring load balance of CombBLAS operations that use the resulting distributed sparse matrix. For this reason alone, one can use the \texttt{ReadGeneralizedTuples} function in lieu of the \texttt{ParallelReadMM} function if the input is known to be severely load imbalanced. In such cases, reading the input into a distributed sparse matrix \texttt{ParallelReadMM} and subsequently permuting it within CombBLAS for load balance might not be feasible, because the load imbalance can be high enough for some process to run out of local memory before \texttt{ParallelReadMM} finishes.

\section{Experimental results}
\label{sec:experiments}
\revision{
Over the past decade, CombBLAS made significant progress in (a) developing new algorithms for sparse-matrix primitives~\cite{azad2016distributed}, (b) implementing algorithms to extract high-performance from heterogeneous distributed systems with CPUs and GPUs~\cite{selvitopi2020optimizing}, (c) demonstrating extreme-scalability using communication-avoiding algorithms that scale to the limit of supercomputers~\cite{batchedSpGEMMIPDPS21, fastsvsiampp20}, and (d) providing customized functionality for several high-impact applications in computational biology~\cite{Azad2018, selvitopi2020distributed} and scientific computing~\cite{azad2020distributed}. } 
While many of these progresses have already been published separately, we show the overall impact of moving CombBLAS 1.0 to CombBLAS 2.0 and demonstrate how CombBLAS 2.0 made important progress toward exascale.

\newcommand{\specialcell}[2][c]{%
  \begin{tabular}[#1]{@{}l@{}}#2\end{tabular}}

\begin{table}[!t]
    \centering
    \caption{Overview of the evaluation platform.}
    \vspace{-4pt}
    \scalebox{0.9}
    {
    \begin{tabular}{r l l l}
    \toprule
    & \textbf{Cori-KNL} & \textbf{Cori-Haswell} & \textbf{Summit} \\
    \toprule
    Processor & \specialcell{Intel Xeon\\Phi 7250} & \specialcell{Intel Xeon\\E5-2698} & IBM Power9 \\
    Accelerator & -- & -- & \specialcell{NVIDIA Volta V100\\(6 per node)} \\
 
    Cores/node & 68 & 32 & 42 \\
    Hyper-threads/core & 4 & 2 & 4 \\
    Memory/node & 112GB & 128GB & \specialcell{512GB Host\\96GB HBM}\\

    \toprule

    Total nodes & 9,668 & 2,388 & 4,608 \\ 
    Total memory & 1.09 PB & 298.5 TB & 10 PB\\
    
    Interconnect & \multicolumn{2}{l}{\specialcell{Cray Aries with \\ Dragonfly topology}} & \specialcell{Non-blocking\\Fat tree} \\
    \toprule
    Compiler &  \multicolumn{2}{l}{\specialcell{Intel icpc 19.0.3\\with -O3}} &  \specialcell{GNU gcc 8.1.1\\with -O3\\NVIDIA nvcc 11.0.3} \\
    MPI Vendor & \multicolumn{2}{l}{Cray-MPICH 7.7.10} & Spectrum MPI 10.3\\
    \bottomrule
    \end{tabular}
    }
    \label{tab:system_info}
\end{table}

\subsection{Experimental settings}
We evaluate the performance of CombBLAS on NERSC Cori and Summit systems. 
In the benchmarking experiment shown in Fig.~\ref{fig:randSpMV}, we used the Cori Haswell partition. 
In the scalability experiments, we use the Cori KNL partition (because the KNL partition has more nodes).
To demonstrate the accelerator support described in Section~\ref{sec:gpus} we use the Summit system.
Table.~\ref{tab:system_info} provides a summary of the systems.
%
On Cori, when the number of thread $t$ is greater than one, we used MPI+OpenMP hybrid parallelization.
When $t$ is equal to one, pure MPI is used.
In either case, only one thread in every process makes MPI calls.

The properties of the matrices used in evaluation are presented in Table~\ref{tab:dataset}.
\revision{Virus, Archaea, Eukarya, Isolates1, and Isolates2 are protein-similarity networks generated from isolate genomes in the IMG
database and are publicly available with HipMCL~\cite{Azad2018}. Metaclust50 stores similarities of proteins in Metaclust50 (https://metaclust.mmseqs.com/) dataset which contains predicted genes from metagenomes and metatranscriptomes of assembled contigs from IMG/M and NCBI. Friendster, it-2004, arabic-2005, and uk-2002 are various social and road networks available in the SuiteSparse Matrix Collection~\cite{Davis2011}.
Finally, the Hyperlink graph is extracted from the web crawl data~\cite{meusel2014graph}.}
For the experiments demonstrating the evolution of SpGEMM performance in CombBLAS over the years (Section~\ref{sec:evol-spgemm}), we use the matrices Eukarya and Friendster.
The experiments demonstrating the scalability of sparse kernels and applications (Section~\ref{sec:scalability}) are conducted on matrices Eukarya, Metaclust50, and Hyperlink.
For the accelerator support demonstration in CombBLAS (Section~\ref{sec:acc-exp}), we use HipMCL~\cite{Azad2018} and PageRank applications.
For the former, we use matrices Archaea, Eukarya, and Isolates1, and for the latter we use matrices it-2004, arabic-2005, and uk-2002.
%


\begin{table}[b]
    \centering
    \caption{Statistics about test matrices used in our  experiments. For SpGEMM experiments, we show $\flop$ and nonzeros in the output. M: million, B: billion and T: trillion.
    }
    \scalebox{0.9}
    {
    \begin{tabular}{l r r r r r}
    \toprule
    Matrix ($\mA$) & rows & columns & $\nnz(\mA)$ & $\nnz(\mC)$ & $\flop$ \\
    \toprule
    Virus & 0.22M & 0.22M & 4.58M & 7.49M & 366M\\
    Archaea & 1.6M & 1.6M & 204M & -- & -- \\
    uk-2002 & 19M & 19M & 298M & -- & -- \\
    Eukarya & 3M & 3M & 360M & 2B & 134B \\
    arabic-2005 & 23M & 23M & 640M & -- & -- \\
    Isolates1 & 8.7M & 8.7M & 1B & 10.7B & 386B \\
    it-2004 & 41M & 41M & 1.2B & -- & -- \\
    Isolates2 & 17.5M & 17.5M & 4.2B & 49B & 3T \\
    Friendster & 66M & 66M & 3.6B & 1T & 1.4T \\
    Metaclust50 & 282M & 282M & 37B & 1T & 92T \\
    Hyperlink & 3.27B & 3.27B & 124.9B & -- & -- \\

    \bottomrule
    \end{tabular}
    }
    
    \label{tab:dataset}
\end{table}

\begin{figure}[t!]
    \centering
    \includegraphics[width=0.48\textwidth]{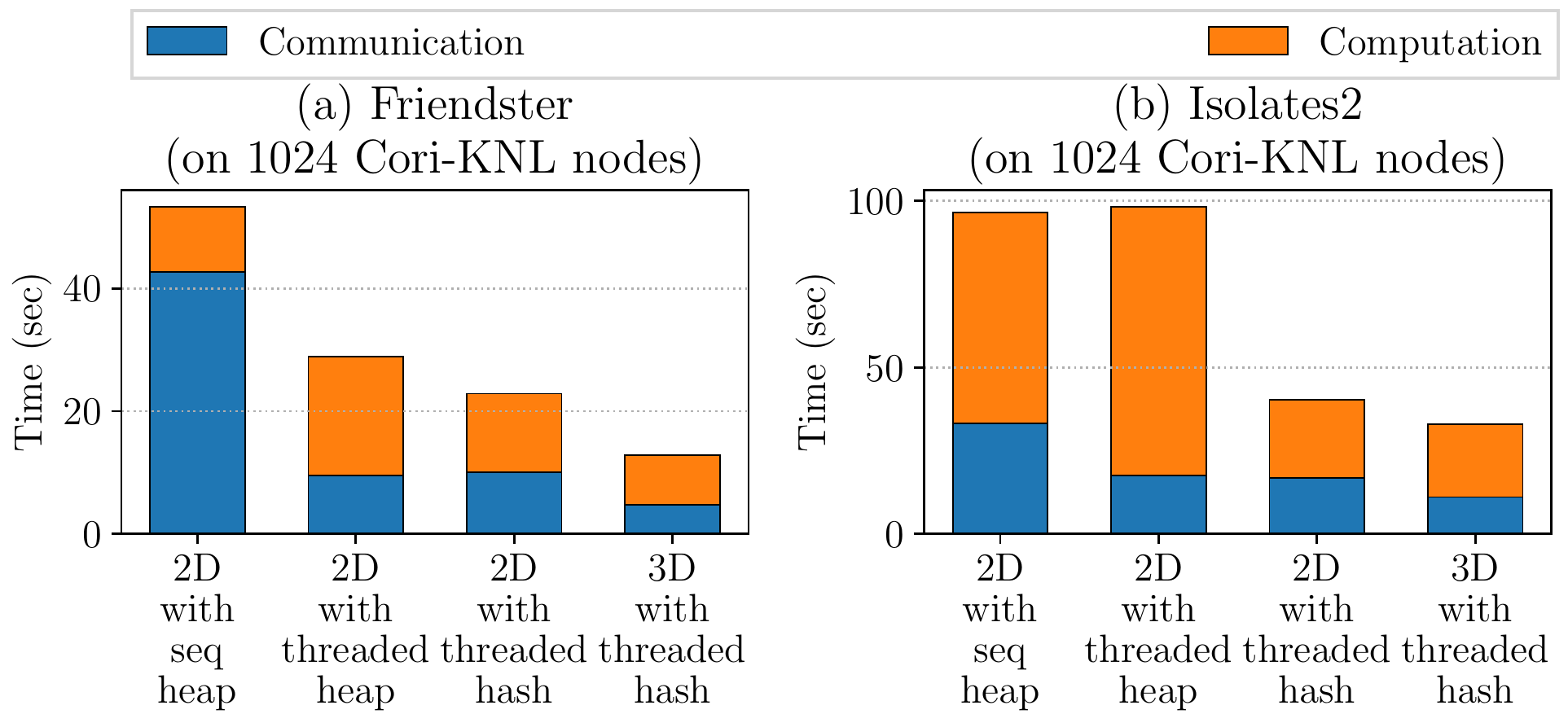}
    \caption{Distributed SpGEMM Evolution. For the Friendster matrix, multiplication was carried on using 4 batches. 64 cores were used for each node. For multithreaded computations, 4 processes per node and 16 threads per process were used. 
    }
    \label{fig:spgemm-evolution}
\end{figure}

\subsection{The evolution of SpGEMM}
\label{sec:evol-spgemm}
SpGEMM is one of the most computationally expensive and widely-used kernels implemented in CombBLAS. 
The original distributed-memory SpGEMM algorithm was based on 2D Sparse SUMMA with a sequential heap-based local SpGEMM implementation~\cite{Buluc2012}. 
With the evolution of computational platforms and application requirements, SpGEMM also evolved by adding new algorithms, implementations, and hardware support. 
First, as supercomputer nodes started to use multicore processors, we implemented multithreaded SpGEMM algorithms.
Second, to ensure extreme scalability, we designed and developed new CA SpGEMM algorithms~\cite{3dspgemmsisc16}. Third, we implemented new hash-based algorithms that are often faster than the heap-based algorithms~\cite{hashspgemmparco19}. Finally, we develop a batched SpGEMM algorithm that forms output in batches when the entire output does not fit in the aggregate memory of all nodes~\cite{batchedSpGEMMIPDPS21}.

Fig.~\ref{fig:spgemm-evolution} compares the computation and communication costs of different variants of SpGEMM. 
We observe that using 16 threads per process in 2D SUMMA, the communication cost decreases in both multiplications. The communication reduction is $4\times$ for Friendster and  $2\times$ for Isolates. 
However, multithreading may increase the computation time because the OpenMP parallel codes do not scale linearly with increased thread counts. By contrast, with perfect load balancing, MPI computations are expected to scale linearly. 
However, with the use of faster processors (e.g., Haswell instead of KNL) and sparse matrices (e.g, Friendster), the communication cost is expected to dominate the computation at extreme scale. 
Hence, multithreading almost always reduces the overall runtime of our distributed SpGEMM. 
Next, the hash SpGEMM algorithm reduces the local computational time significantly for both matrices in Fig.~\ref{fig:spgemm-evolution}.
The CA SpGEMM further reduces the communication time shown in the last bar.
Thus, using the same computational resources, new algorithms and implementations of SpGEMM in CombBLAS 2.0 run $3\times$ to $5\times$ faster than equivalent algorithms in CombBLAS 1.0.
Furthermore, the entire output from squaring the Friendster matrix cannot even be stored on 1024 nodes because of memory limitations (Table~\ref{tab:dataset} shows the output of this multiplication has 1 trillion nonzeros). 
CombBLAS enables such multiplication by generating results in batches; for example, 4 batches are used in Fig.~\ref{fig:spgemm-evolution}(a).
This SpGEMM algorithm can be directly used with applications that do not access the whole output matrix all at once, such as many-to-many protein sequence aligner PASTIS~\cite{selvitopi2020distributed} and protein clustering algorithm HipMCL~\cite{Azad2018}.

\begin{figure}[!t]
    \centering
    \includegraphics[width=0.37\textwidth]{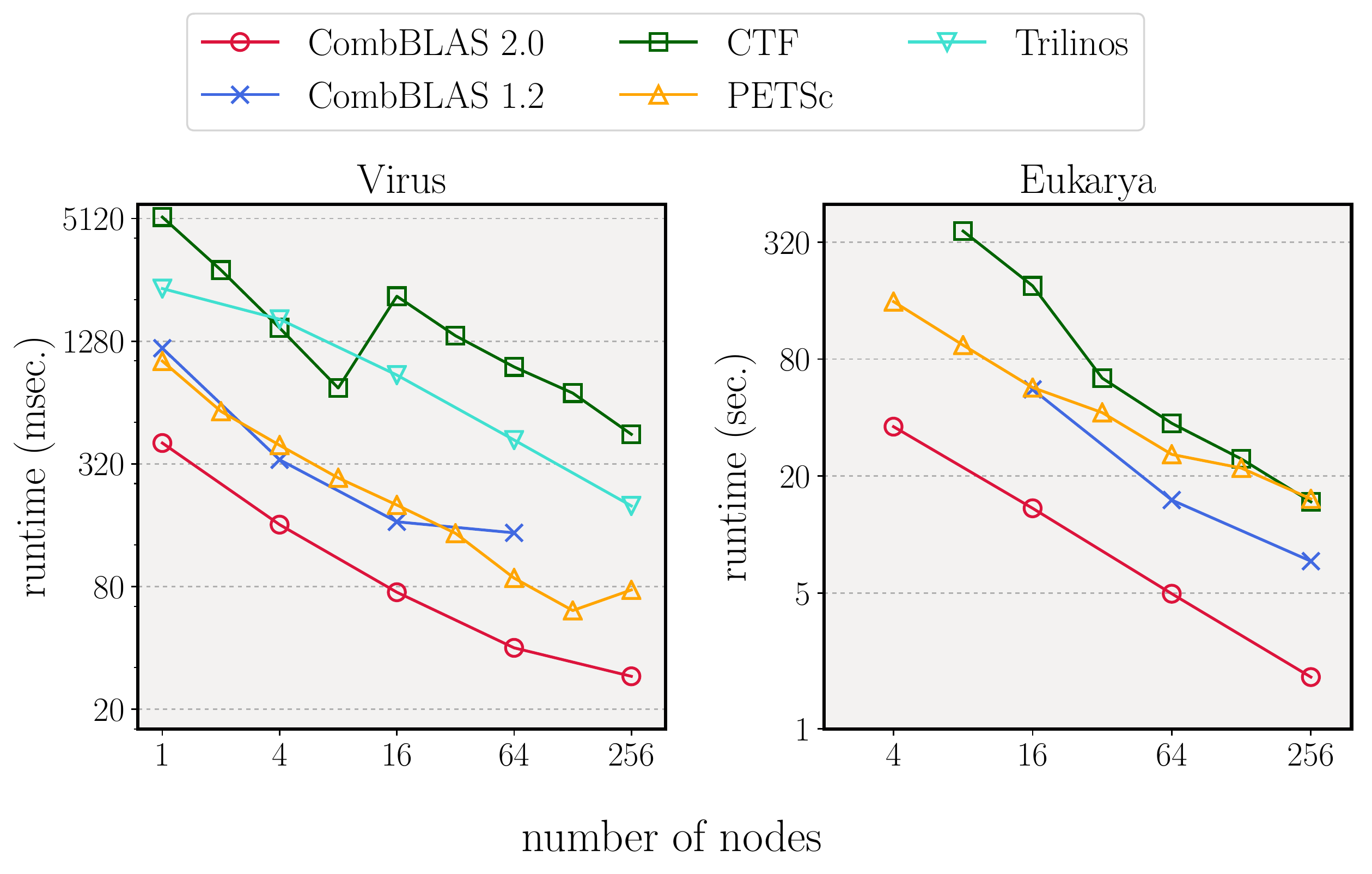}
    \caption{\revision{Parallel SpGEMM runtime of two versions of CombBLAS and other popular parallel sparse linear algebra libraries on two different matrices.}  
    }
    \label{fig:combblas-vs-others}
\end{figure}

\subsection{Comparison against other parallel sparse linear algebra libraries}
\label{sec:combblas-comparison}
\revision{
We assess the parallel runtime and I/O performance of CombBLAS against three popular parallel sparse linear algebra libraries,  CTF~\cite{solomonik2015sparse}, PETSc~\cite{Petsc2014}, and Trilinos~\cite{Heroux2005}. 
While these libraries have different focuses, they all support similar parallel sparse matrix operations.
We also evaluate CombBLAS 2.0 against the earlier version 1.2.
We specifically focus on the performance of SpGEMM and measure its parallel runtime for all evaluated libraries.
For these experiments, we use the Virus and Eukarya networks, and run on the Haswell partition of Cori with varying numbers of nodes from 1 to 256.
%
%
We use 4 MPI processes per node and 8 threads per process for all libraries except CombBLAS 1.2, which did not have multithreaded SpGEMM.
For CombBLAS 1.2, we test out various number of MPI tasks per node and report the result of the best run.
We plot the obtained parallel runtime results in Fig.~\ref{fig:combblas-vs-others} and the I/O times in Table~\ref{tab:io}.
%
%
%
Because Trilinos takes too long to read its input matrices, we exclude it from our evaluations for the larger Eukarya network.

Fig.~\ref{fig:combblas-vs-others} shows that the SpGEMM operation in CombBLAS 2.0 is both more scalable and 3x-4x faster compared to the implementation in CombBLAS 1.2.
%
%
Additionally, CombBLAS 2.0 is able to run on instances where CombBLAS 1.2 fails due to being out of memory.

\begin{table}[h]
\begin{center}
\caption{I/O time (seconds) of CombBLAS and various parallel sparse linear algebra libraries for two networks.}
\scalebox{0.8}
{
  \begin{tabular}{l r r r r r r}
  \toprule
  & & \multicolumn{2}{c}{CombBLAS} &  &  &  \\
  \cmidrule{3-4}
  Matrix & nodes & ASCII & binary & CTF & PETSc & Trilinos \\
  \midrule
  \multirow{3}{*}{Virus} &  4 & 0.83 & 0.40 & 0.75 & 0.34 & 145.37 \\
  &                        16 & 0.41 & 0.32 & 0.54 & 0.31 & 190.54 \\
  &                        64 & 0.29 & 0.48 & 4.69 & 0.61 & 895.33 \\
  \midrule
   \multirow{3}{*}{Eukarya} &  16 & 19.41   & 15.70 & 43.60 & 13.33 & -- \\
  &                        64 & 5.31 & 6.95 & 26.38 & 9.91 & -- \\
  &                        256 & 2.31 & 5.74 & 26.76 & 9.52 & -- \\
  \bottomrule
  \end{tabular}%
}
\label{tab:io}%
\end{center}
\end{table}

CombBLAS 2.0 is faster and more scalable than the three other libraries evaluated for the SpGEMM operation. 
%
%
Among the evaluated libraries, the closest competitor is PETSc, but CombBLAS 2.0 is still 3x-8x faster and more scalable than PETSc.
PETSc's performance is similar to that of CombBLAS 1.2, and the other two libraries, CTF and Trilinos trail PETSc and CombBLAS 1.2.
%

\begin{figure}[!t]
    \centering
    \includegraphics[width=0.48\textwidth]{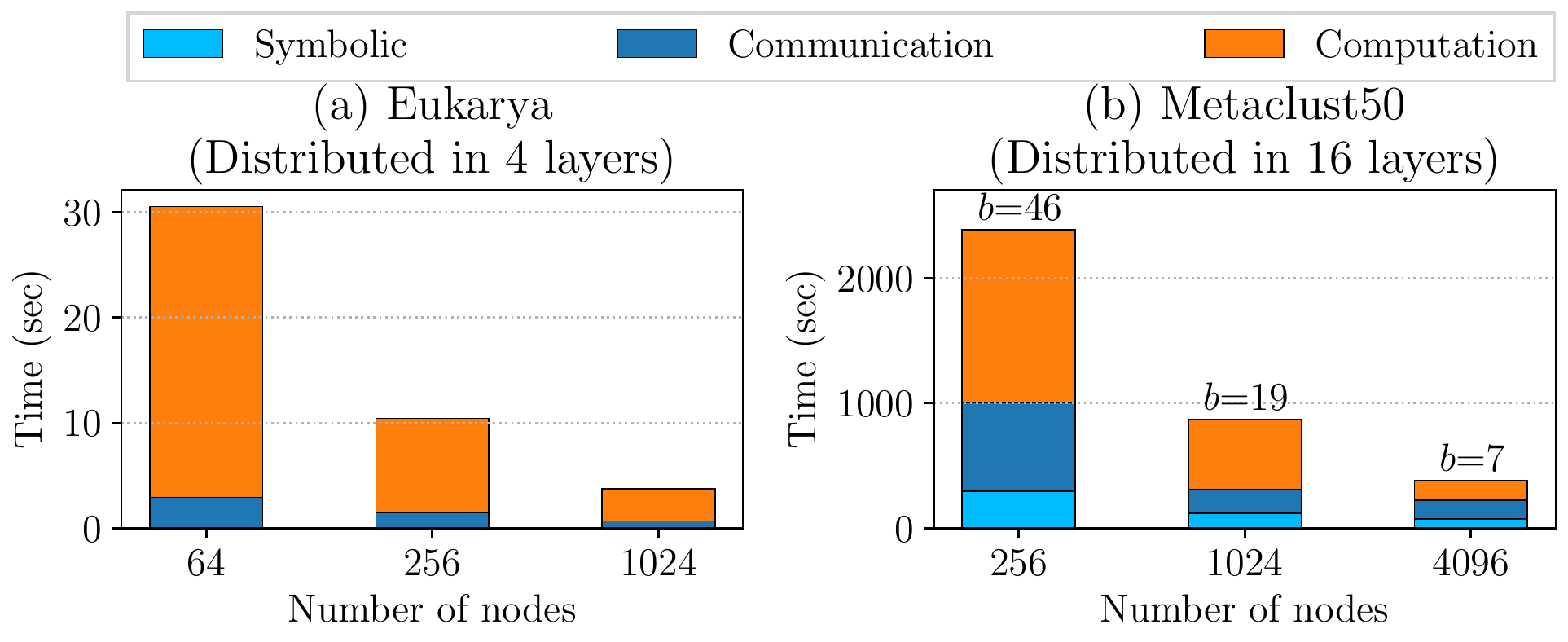}
    \caption{Strong scaling of CA SpGEMM when squaring (a) Eukarya and (b) Metaclust50.  
    We used 4 MPI processes per node and 16 threads per process on Cori KNL. Number of layers is mentioned at the top of each figure. For Metaclust50, $b$ on top of each bar represents the number of batches needed for SpGEMM (because of the memory limitation). $b$ is calculated using a symbolic step. No batching is needed for Eukarya. 
    }
    \label{fig:spgemm-scaling}
\end{figure}

Table~\ref{tab:io} shows that CombBLAS and PETSc have faster and more scalable I/O compared to Trilinos and CTF.
PETSc requires a custom binary file format for its input, while CTF requires a custom ASCII text format, for which we needed to convert the matrices explicitly.
CombBLAS and Trilinos support the common Matrix Market format.
In addition, CombBLAS supports a custom binary format, and the I/O performance of CombBLAS in both ASCII and binary formats are fast enough to be of practical use.
}

\subsection{Extreme scalability of operations and applications}
\label{sec:scalability}
Thanks to hybrid parallelization techniques and CA algorithms, CombBLAS 2.0 successfully scales to thousands of nodes on modern supercomputers. 
Here we demonstrate the strong scaling of CA SpGEMM and a connected component algorithm based on SpMV and SpMSpV.

Fig.~\ref{fig:spgemm-scaling} shows the strong scaling of CA SpGEMM when squaring a small-scale (Eukarya) and a large-scale (Metaclust50) matrix. 
Both matrices represent protein similarity networks, and squaring these matrices captures random walks on a graph needed by the protein clustering application HipMCL~\cite{Azad2018}.
Since Eukarya is a small-scale matrix with the runtime dominated by computation, we used just 4 layers in the CA SpGEMM algorithm.
For large-scale experiments with Metaclust50, we used  16 layers.
When we go from 64 nodes to 1024 nodes (a $16\times$ increase in node counts), CA SpGEMM with Eukarya runs $7\times$ faster. 
Similarly, when we go from 256 nodes to 4096 nodes (a $16\times$ increase in node counts), CA SpGEMM with Metaclust50 becomes $7\times$ faster. 
Hence, CA SpGEMM works equally well for both small-scale and large-scale matrices on different concurrencies. 
More importantly, the communication cost continues to reduce with increasing node counts.
Thus, given big enough matrices, distributed SpGEMM in CombBLAS 2.0 has the ability to scale to the limit of modern supercomputers. 

\begin{figure}[!t]
    \centering
    \includegraphics[width=\linewidth]{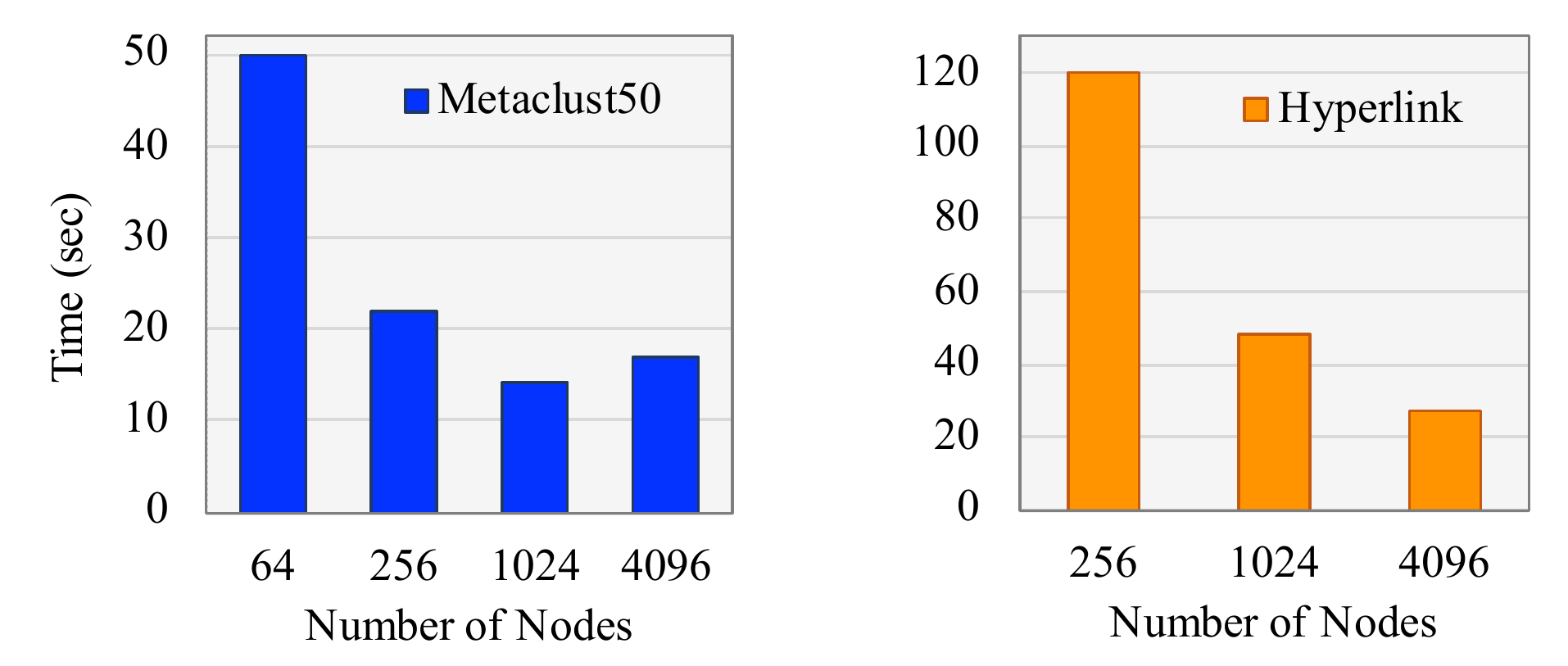}
    \caption{Scaling FastSV (a distributed-memory connected-component algorithm)  on Cori KNL.  We used 4 MPI processes per node and 16 threads per process. 
    }
    \label{fig:FastSV-scaling}
\end{figure}

Fig.~\ref{fig:FastSV-scaling} shows the strong scaling of FastSV~\cite{fastsvsiampp20}, an algorithm for finding connected components in an undirected graph. 
FastSV is based on the classic Shiloach-Vishkin algorithm~\cite{shiloach1980log} and developed using SpMV, SpMSpV and vector assign/extract operations from CombBLAS.
Despite relying on matrix-vector multiplications and vector manipulations, FastSV scales remarkably well up to thousands of nodes as shown in Fig.~\ref{fig:FastSV-scaling}.
For the relatively small matrix Metaclust50, FastSV does not scale beyond 1024 nodes because of the scaling limitation of SpMSpV. 
For the Hyperlink graph with 124.9 billion edges, FastSV scales to 4096 nodes (262,144 cores).
Since these graphs require more than 1TB
memory just to store the input, it is often impossible to analyze them on a typical shared-memory server.
Thus, CombBLAS 2.0 expands the scalability horizon of graph analysis, even for algorithms with relatively low computational complexities and irregular communication patterns.

\subsection{Benefiting from accelerators}
\label{sec:acc-exp}
%
In this section, we demonstrate two applications powered by CombBLAS partially running on NVIDIA GPUs: High Performance Markov Clustering (HipMCL) optimized for systems containing GPU nodes~\cite{selvitopi2020optimizing} and the PageRank algorithm.
The former makes heavy use of the SpGEMM kernel while the latter mainly relies on SpMV.
We use the Summit system for both of these applications.
We use one MPI task for each GPU, which results in 6 MPI tasks per node.
We use rather modest-sized matrices for these applications -- the largest ones containing around 1 billion nonzeros (Table~\ref{tab:dataset}).
%

\begin{figure}[t!]
    \centering
    \begin{subfigure}[t]{0.47\textwidth}
        \centering
        \includegraphics[width=1.0\textwidth]{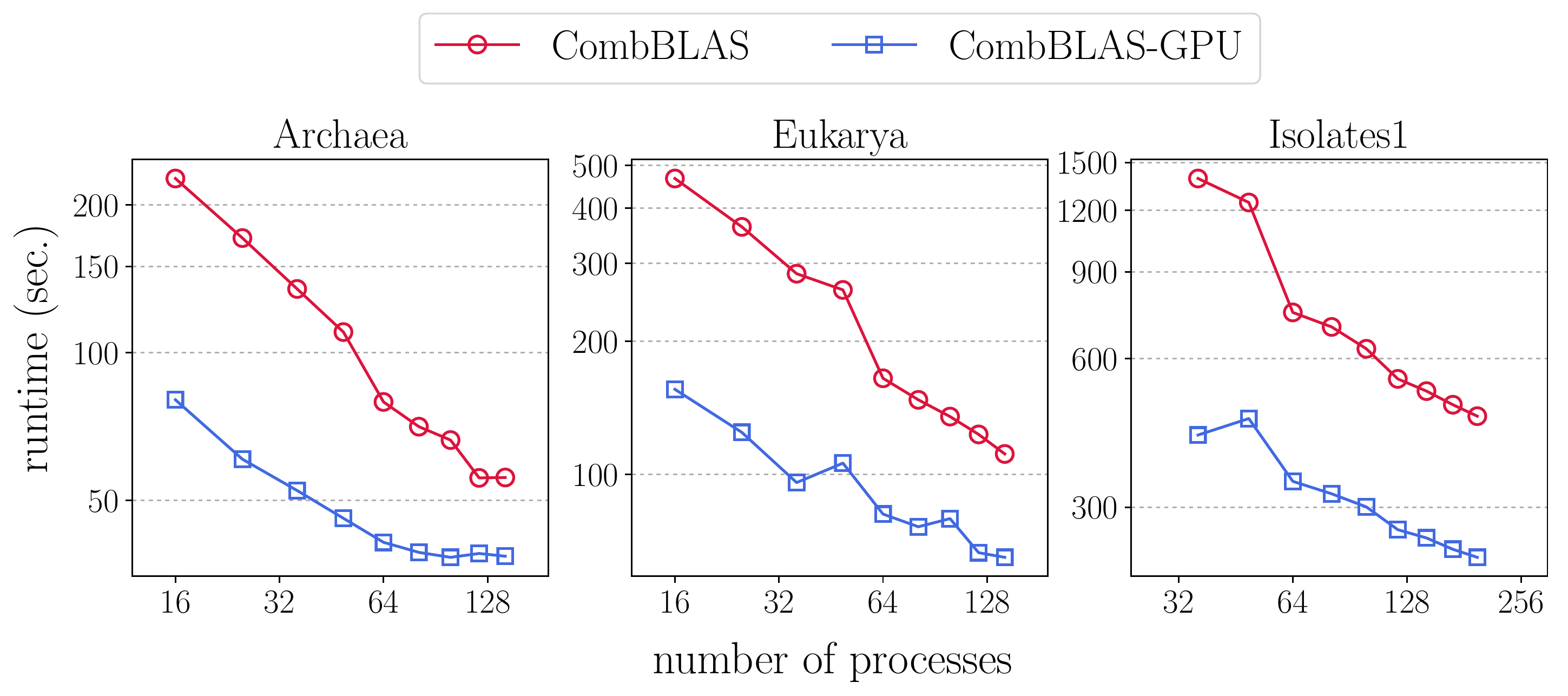}
        \caption{Scalability of CombBLAS with and without GPU support.}
        \label{fig:hipmcl-cpu-vs-gpu}
    \end{subfigure}
    \par\bigskip 
    \begin{subfigure}[t]{0.47\textwidth}
        \centering
        \includegraphics[width=1.0\textwidth]{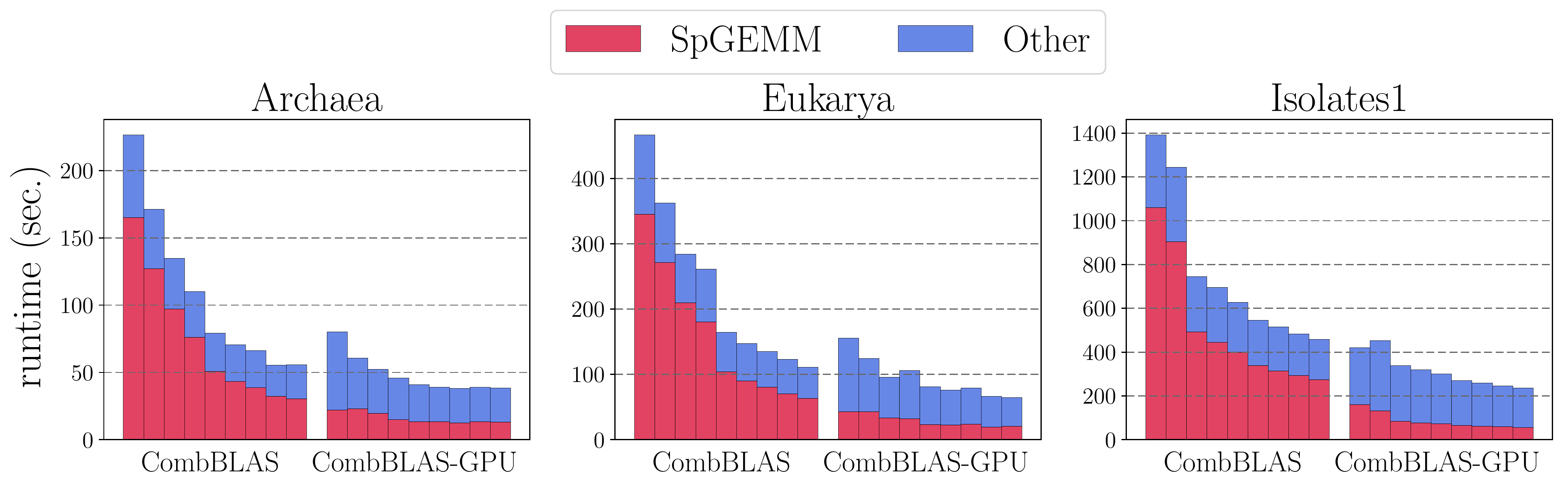}
        \caption{Time spent in SpGEMM.}
        \label{fig:hipmcl-spgemm-bars}
    \end{subfigure}
    \caption{HipMCL with and without accelerator support. Number of processes $\{p=q^2: q=4,5,\ldots,12\}$ for Archaea and Eukarya, and $\{p=q^2: q=6,7,\ldots,14\}$ for Isolates1. 
    }
    \label{fig:hipmcl-gpu}
\end{figure}

Figure~\ref{fig:hipmcl-gpu} shows HipMCL running with and without accelerator support.
SpGEMM constitutes a very significant portion of overall execution time as seen in Figure~\ref{fig:hipmcl-spgemm-bars}.
By offloading these computationally expensive SpGEMM operations to GPUs, CombBLAS is able to obtain $2{-}3\times$ speedup on all three tested instances.
The GPUs are especially effective when the compression ratio of the multiplied matrices is high -- which is usually the case for these instances.
Note that the baseline CPU-based HipMCL here uses the fastest multithreaded multiplication algorithms presented in this work.
GPU support in CombBLAS 2.0 provides more than an order of magnitude speedup~\cite{selvitopi2020optimizing}.

\begin{figure}[t!]
    \centering
    \begin{subfigure}[t]{0.47\textwidth}
        \centering
        \includegraphics[width=1.0\textwidth]{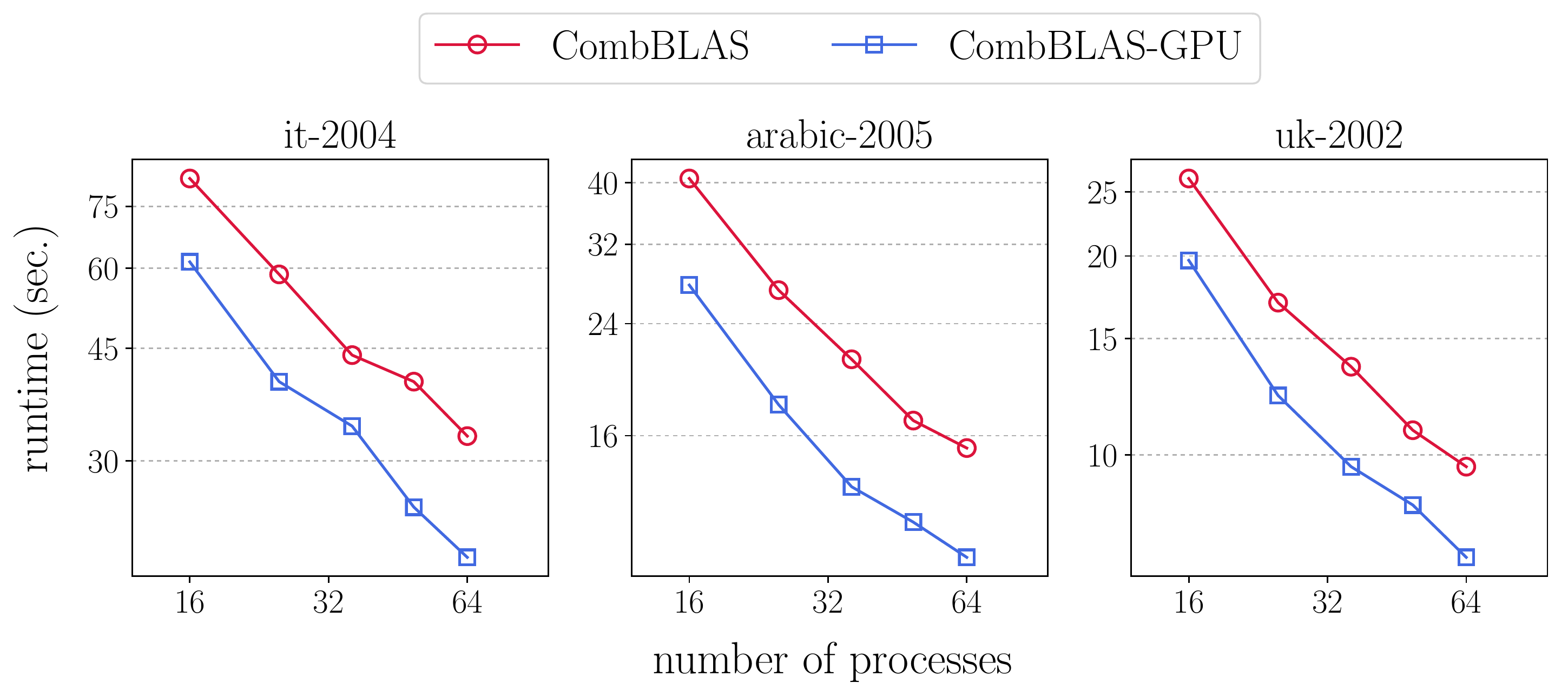}
        \caption{Scalability of CombBLAS with and without GPU support.}
        \label{fig:pagerank-cpu-vs-gpu}
    \end{subfigure}
    \par\bigskip 
    \begin{subfigure}[t]{0.47\textwidth}
        \centering
        \includegraphics[width=1.0\textwidth]{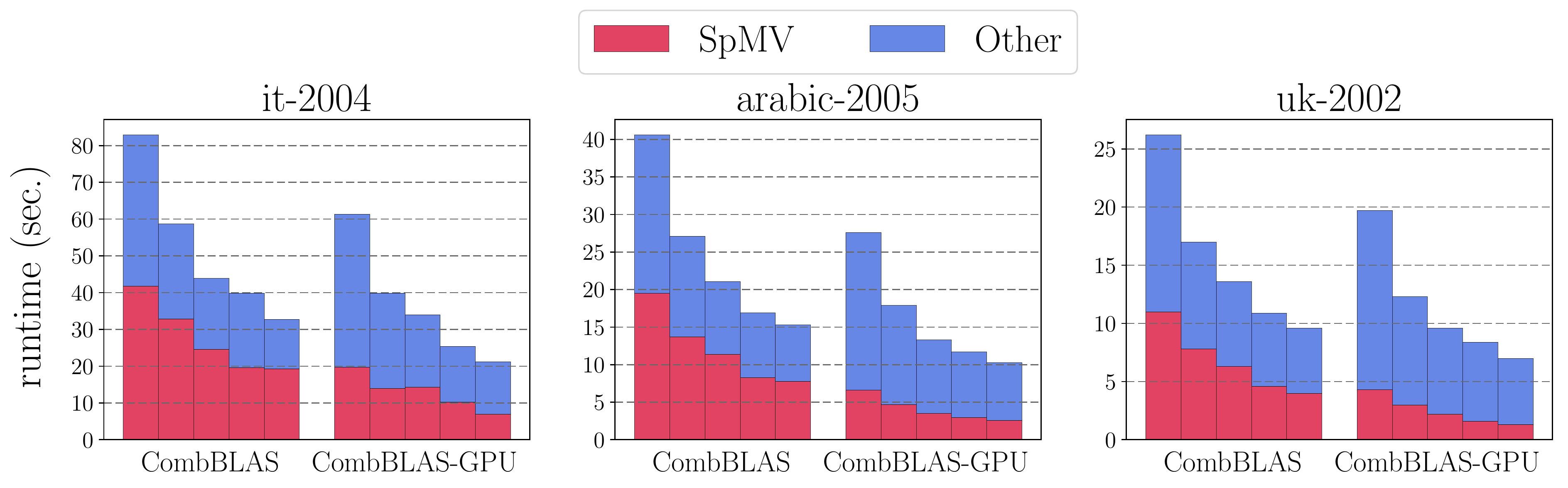}
        \caption{Time spent in SpMV.}
        \label{fig:pagerank-spmv-bars}
    \end{subfigure}
    \caption{PageRank with and without accelerator support. Number of processes $\{p=q^2: q=4,5,\ldots,8\}$ for all three graphs. 
    }
    \label{fig:pagerank-gpu}
\end{figure}

The results for PageRank are presented in Figure~\ref{fig:pagerank-gpu}.
SpMV has lower arithmetic intensity than SpGEMM, and this is reflected in Figure~\ref{fig:pagerank-cpu-vs-gpu} where the performance improvement due to GPU utilization is around $20{-}40\%$, which is smaller than the improvement observed in Figure~\ref{fig:hipmcl-cpu-vs-gpu}.
Although we utilize the algorithms recommended by NVIDIA for SpMV in cuSPARSE, the input matrix needs to be transposed due to lack of CSC matrix storage support in cuSPARSE.
Despite this slowdown, utilizing GPUs still results in considerable performance gain.
%


\section{Conclusions}
\label{sec:conclusions}

CombBLAS is currently the primary backend of several distributed-memory applications such as the graph clustering code HipMCL~\cite{Azad2018}, protein homology search code PASTIS~\cite{selvitopi2020distributed}, and read-to-read genome aligner diBELLA.2D~\cite{guidi2020parallel}. Several other applications such as the sparse direct solver SuperLU~\cite{li2003superlu_dist}, the structured dense matrix solver STRUMPACK~\cite{rouet2016distributed}, and the sketching library libskylark~\cite{libskylark} also utilize CombBLAS for certain tasks.

As evidenced by the experiments shown in this paper, CombBLAS is able to work on sparse matrix operations where the output has $1$ trillion nonzeros. Contemporary publications that use CombBLAS also demonstrated that it can successfully utilize $1$ million hardware threads~\cite{batchedSpGEMMIPDPS21}.

\section*{Acknowledgments}
 We would like to acknowledge the assistance of Penporn Koanantakool and Alok Tripathy with some of the figures. We thank Scott Beamer, Saliya Ekanayake, Giulia Guidi, Tristan Konolige, Adam Lugowski, and Alex Reinking for their contributions on various issues related to CombBLAS development or installation. 
 
  This work is supported by the Advanced Scientific Computing Research (ASCR) program within the Office of Science of the DOE under contract number DE-AC02-05CH11231, and by the National Science Foundation 
under Award No. 1823034. This research was also supported by the Exascale Computing Project (17-SC-20-SC), a collaborative effort of the U.S. Department of Energy Office of Science and the National Nuclear Security Administration. John Gilbert is partially supported by NSF Grant CCF-1637564.

This research used resources of the Oak Ridge Leadership Computing Facility at
the Oak Ridge National Laboratory, which is supported by the Office of Science
of the U.S. Department of Energy under Contract No. DE-AC05-00OR22725.

\bibliographystyle{IEEEtran}
\bibliography{3DSpGEMM, references}

%

\vspace{-5 mm}
\begin{IEEEbiography}[{\includegraphics[width=1in,height=1.25in,clip,keepaspectratio]{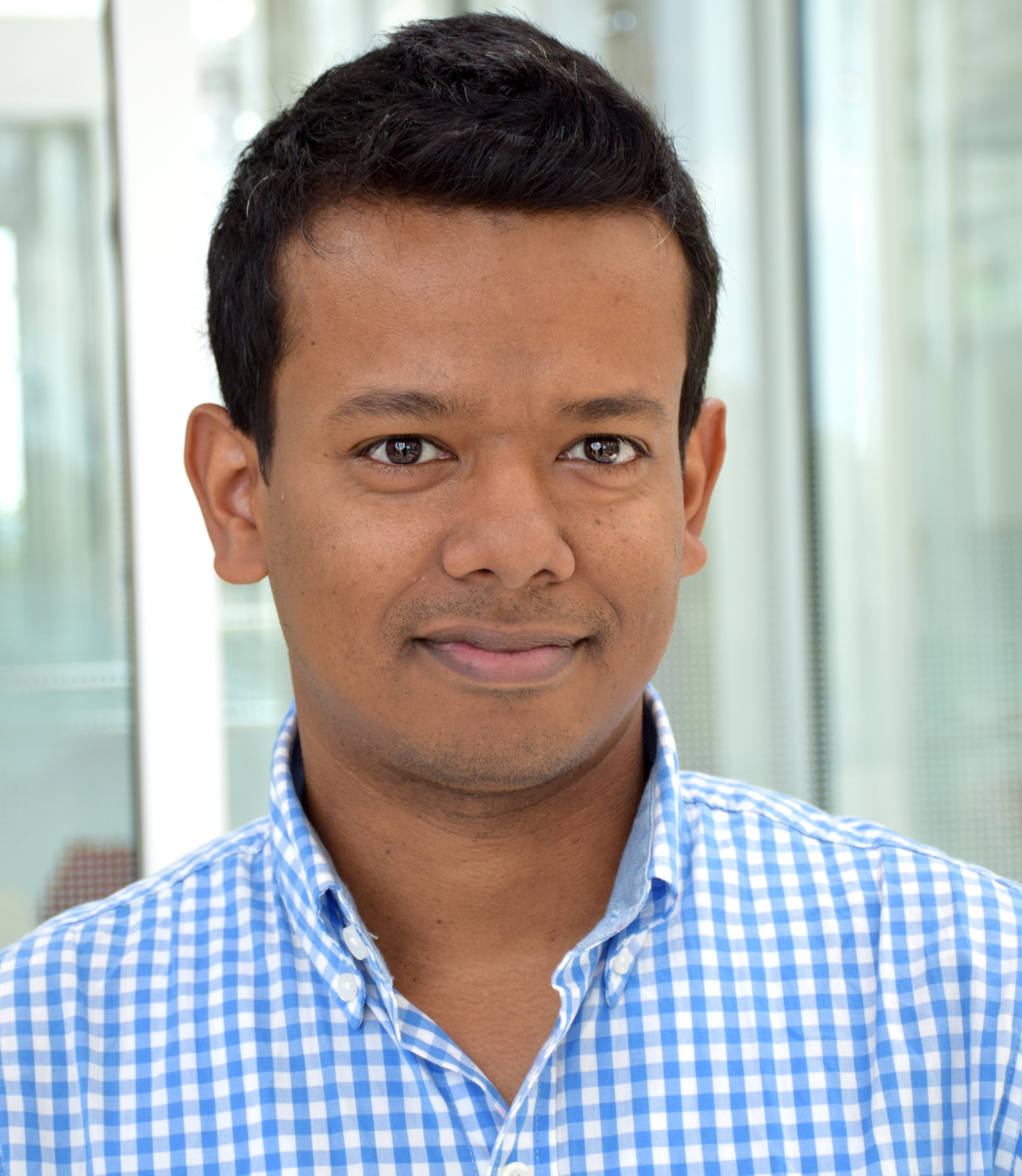}}]{Ariful Azad}
is an Assistant Professor of Intelligent Systems Engineering at Indiana University. He got his PhD from Purdue and then worked as a Research Scientist at Lawrence Berkeley National Laboratory. His research interests are in high-performance graph analysis and learning.
\end{IEEEbiography}

\vspace{-5 mm}
\begin{IEEEbiography}[{\includegraphics[width=1in,height=1.25in,clip,keepaspectratio]{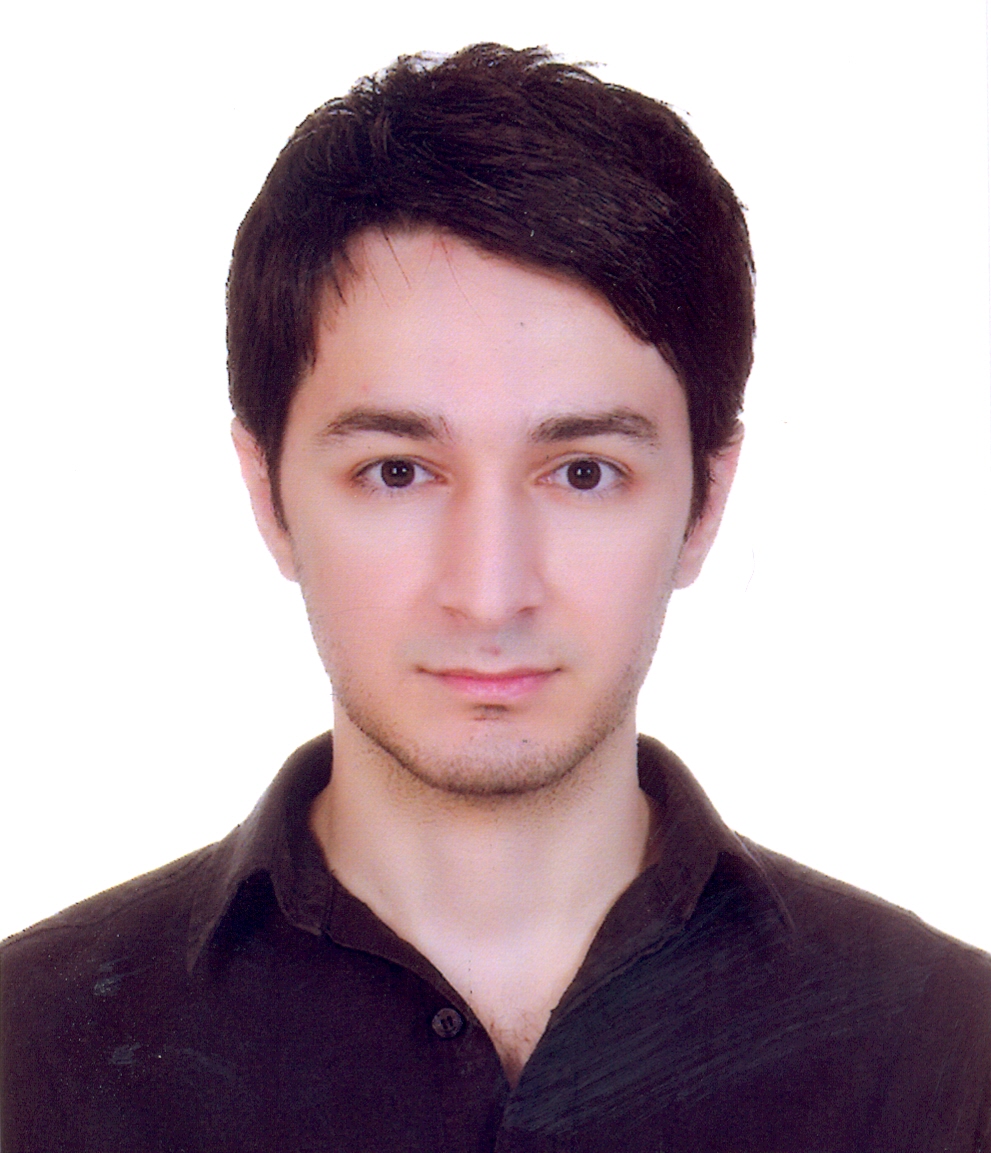}}]{Oguz Selvitopi}
got his PhD from Bilkent University in 2016 and he is currently a research scientist at Lawrence Berkeley National Laboratory. His research interests include high performance computing, parallel graph algorithms, and bioinformatics.
\end{IEEEbiography}

\vspace{-5 mm}
\begin{IEEEbiography}[{\includegraphics[width=1in,height=1.25in,clip,keepaspectratio]{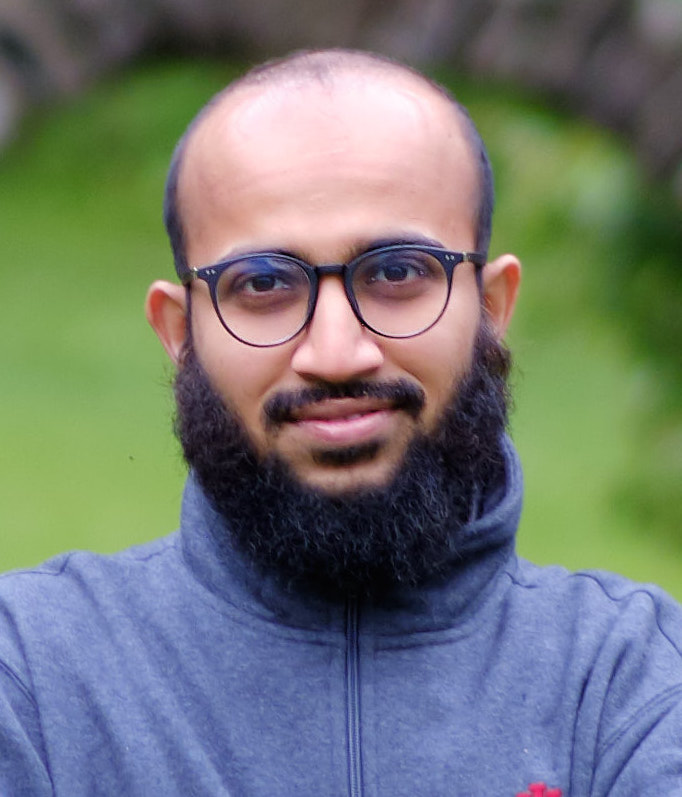}}]{Md Taufique Hussain}
is a PhD student at Indiana University Bloomington. He is studying high performance graph algorithms.
\end{IEEEbiography}

\vspace{-5 mm}
\begin{IEEEbiography}[{\includegraphics[width=1in,height=1.25in,clip,keepaspectratio]{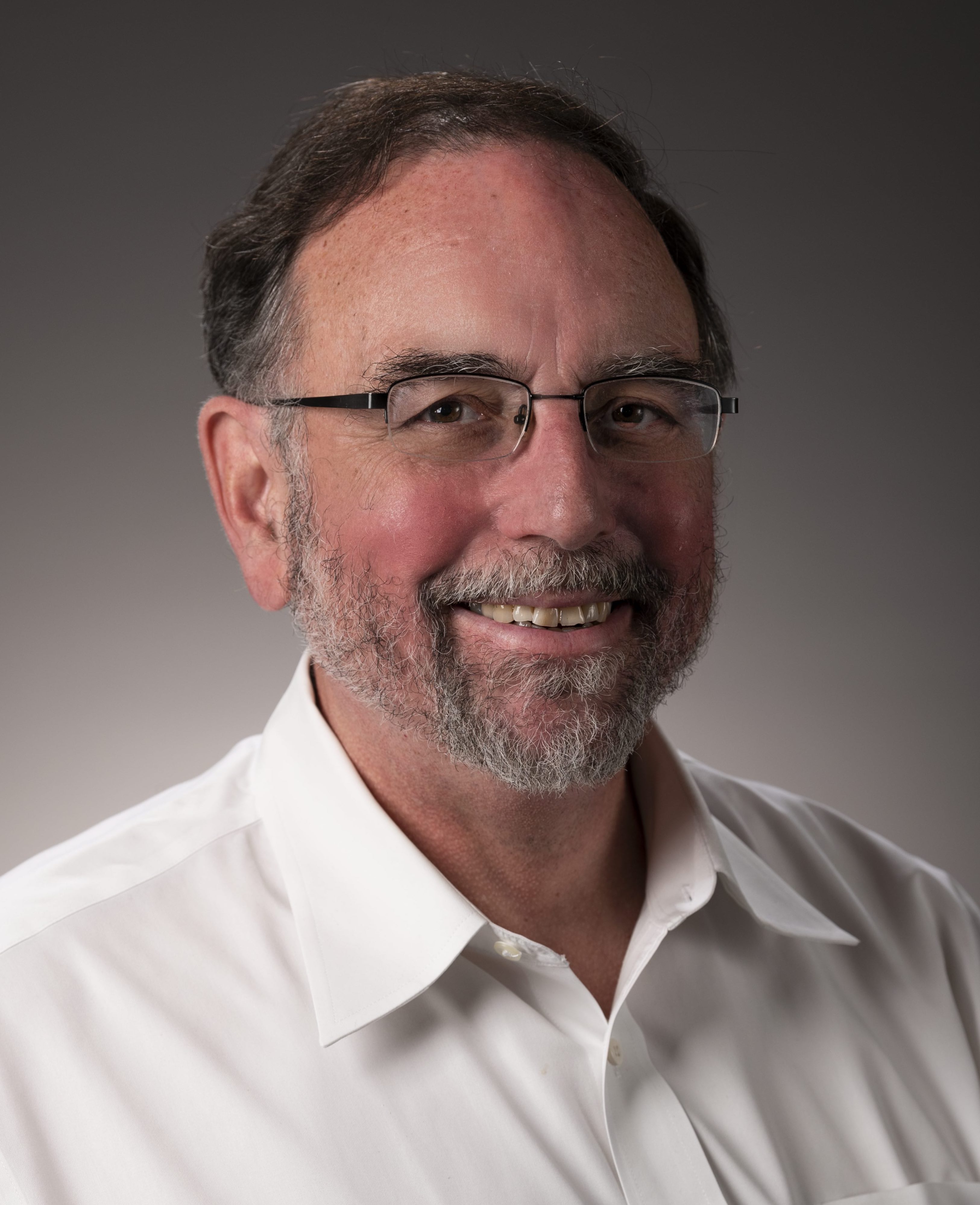}}]{John R. Gilbert}
studies graphs, matrices, and high-performance computing. He got his PhD from Stanford in 1981 and then worked at Cornell and Xerox PARC. He is currently Professor of Computer Science at the University of California, Santa Barbara and Fellow of the Society for Industrial and Applied Mathematics.
\end{IEEEbiography}

\vspace{-5 mm}
\begin{IEEEbiography}[{\includegraphics[width=1in,height=1.25in,clip,keepaspectratio]{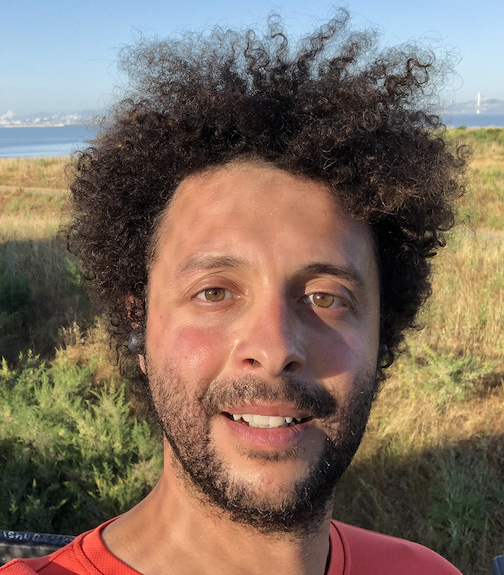}}]{Ayd\i n~Bulu\c{c}}
is a Staff Scientist at the Lawrence Berkeley National Laboratory and an Adjunct Assistant Professor of Electrical Engineering and Computer Sciences at the University of California, Berkeley. His research interests include parallel computing, combinatorial scientific computing, high-performance graph analysis and machine learning, sparse matrix computations, and computational biology.  
\end{IEEEbiography}





\end{document}